\documentclass[lettersize,journal]{IEEEtran}
\pdfoutput=1
\newcommand{\columnversion}[2]{#2}

\usepackage{amsmath,amsfonts}
\usepackage{algorithmic}
\usepackage{algorithm}
\usepackage{array}
\usepackage[caption=false,font=normalsize,labelfont=sf,textfont=sf]{subfig}
\usepackage{textcomp}
\usepackage{stfloats}
\usepackage{url}
\usepackage{verbatim}
\usepackage{graphicx}
\usepackage{cite}
\usepackage[utf8]{inputenc}
\usepackage{array}  
\usepackage{graphicx}
\usepackage{latexsym}
\usepackage{amssymb}
\usepackage{color}
\usepackage{multirow}
\usepackage{algorithmic,algorithm}
\usepackage{comment}
\usepackage{cuted}
\usepackage{mathtools}
\usepackage{footnote}
\usepackage{subfig}
\usepackage{makecell}
\usepackage{pifont}

\newtheorem{lemma}{\bf{Lemma}}
\newtheorem{proposition}{\bf{Proposition}}

\newtheorem{corollary}{\bf{Corollary}}

\def\Nt{N_t}
\def\Tr{{\rm Tr}}
\def\rmH{{\rm H}}
\def\rmT{{\rm T}}
\def\maB{\mathbf{B}}
\def\maI{\mathbf{I}}
\def\maU{\mathbf{U}}
\def\maW{\mathbf{W}}
\def\vew{\mathbf{w}}

\def\maC{\mathbf{C}}
\def\maR{\mathbf{R}}

\def\maPi{\mathbf{\Pi}}
\def\maSigma{\mathbf{\Sigma}}
\def\maLambda{\mathbf{\Lambda}}
\def\hath{\hat{\mathbf{h}}}

\def\maV{\mathbf{V}}
\def\maU{\mathbf{U}}
\def\maH{\mathbf{H}}
\def\maI{\mathbf{I}}
\def\maX{\mathbf{X}}

\def\hatSigma{\hat{\mathbf{\Sigma}}}
\def\hatLambda{\hat{\mathbf{\Lambda}}}
\def\vech{\mathbf{h}}
\def\vecz{\mathbf{z}}
\def\vecx{\mathbf{x}}
\def\vecv{\mathbf{v}}
\def\vecw{\mathbf{w}}
\def\vecc{\mathbf{c}}
\def\vecb{\mathbf{b}}

\def\vecu{\mathbf{u}}
\def\hatc{\hat{\mathbf{c}}}
\def\tilh{\tilde{\mathbf{h}}}

\def\hatb{\hat{\mathbf{b}}}
\def\vecc{\mathbf{c}}

\def\be{\begin{equation}}
\def\ee{\end{equation}}

\def\calH{{\mathcal{H}}}
\def\calG{{\cal G}}
\def\calC{{\cal C}}
\def\calB{{\cal B}}

\def\calV{{\cal V}}
\def\calW{{\cal W}}

\def\vecb{\mathbf{b}}
\def\veca{\mathbf{a}}
\def\bea{\begin{eqnarray}}
\def\eea{\end{eqnarray}}

\DeclarePairedDelimiter{\diagfences}{(}{)}
\newcommand{\diag}{\operatorname{diag}\diagfences}

\newcommand{\mat}[2]{\left[\begin{array}{#1} #2 \end{array}\right]}

\def\eps{\epsilon}
\def\vece{\mathbf{e}}
\def\CC{\mathbb{C}}
\def\vecq{\mathbf{q}}
\def\Voronoi{V}
\newcommand{\mabel}[1]{\label{#1}}

\newcommand{\red}[1]{{\color{red} #1}}

\def\matH{\mathbf{H}}
\def\matZ{\mathbf{Z}}
\def\matA{\mathbf{A}}

\def\vecn{\mathbf{n}}
\def\vecy{\mathbf{y}}

\def\|{|}

\begin{document}
\title{Modular CSI Quantization for FDD Massive MIMO Communication} 

\author{Jialing Liao,~\IEEEmembership{Member,~IEEE}, Roope
  Vehkalahti,~\IEEEmembership{Member,~IEEE},
  Tefjol Pllaha,\\  Wei Han, 
Olav Tirkkonen,~\IEEEmembership{Fellow,~IEEE}     
\thanks{J. Liao  was with the Department of Communications and Networking (Comnet), Aalto University, Finland. She now is with the Department of Electrical Engineering (ISY), Linköping University, Sweden (e-mail: $\rm  jialing.liao@ieee.org$).  
  Roope Vehkalahti was with Comnet, Aalto University, Finland. He now is with the Department of Mathematics and
  Statistics, University of Jyväskylä, Finland (e-mail: $\rm
  roope.i.vehkalahti@jyu.fi$).
  Tefjol Pllaha was with Comnet, Aalto University, Finland. He now is with the Department of Mathematics, University
  of Nebraska–Lincoln, United States (e-mail: $\rm
  tefjol.pllaha@unl.edu$).
  Wei Han is with Huawei Technologies Co., Ltd., Shanghai, P. R.
  China, (e-mail: $\rm wayne.hanwei@huawei.com$).
O. Tirkkonen is with Comnet, Aalto University, Finland (e-mail: $\rm  olav.tirkkonen@aalto.fi$).
This work was funded in part by Huawei Technologies Co., Ltd. 
Part of the paper has  been published in VTC Spring 2021 \cite{Vehkalahti2021}.}}

\markboth{Journal of \LaTeX\ Class Files,~Vol.~14, No.~8, August~2021}%
{Shell \MakeLowercase{\textit{et al.}}: A Sample Article Using IEEEtran.cls for IEEE Journals}


\maketitle

\begin{abstract}
We consider high-dimensional MIMO transmissions in frequency division
duplexing (FDD) systems. For precoding, the frequency selective
channel has to be measured, quantized and fed back to the base station
by the users. When the number of antennas is very high this typically
leads to prohibitively high quantization complexity and large
feedback. In 5G New Radio (NR), a modular quantization approach has
been applied for this, where first a low-dimensional subspace is
identified for the whole frequency selective channel, and then subband
channels are linearly mapped to this subspace and quantized. We
analyze how the components in such a modular scheme contribute to the
overall quantization distortion. Based on this analysis we improve the
technology components in the modular approach and propose 
an orthonormalized wideband precoding scheme and a sequential wideband
precoding approach which provide considerable gains over the
conventional method.
We compare the performance of the developed quantization schemes to
prior art by simulations in terms of the projection distortion, overall
distortion and spectral efficiency, in a scenario with a realistic
spatial channel model.
\end{abstract}


\section{Introduction}\mabel{sec:intro}
\IEEEPARstart{M}{assive} MIMO (mMIMO)
communication~\cite{Marzetta2013} with very large antenna arrays at
the base station (BS) is one of the key components in the 5G New Radio
(NR). Using mMIMO in downlink, however, relies on the availability of
high quality channel state information (CSI) at the transmitter (Tx). In a
Frequency Division Duplex (FDD) system the channel has to be measured
by the users, quantized, and then fed back to the BS. FDD MIMO finite
feedback principles are well understood both in the single user \cite{Hamalainen2000,Love2003,Dai2008} and
multiuser~\cite{Jindal2006,Lee2008,Caire2013} context. With a high number
$N_t$ of BS antennas, however, quantization complexity and the amount
of feedback can be prohibitively high. This problem becomes
particularly difficult in a multiuser-MIMO setting, where the amount
of feedback should scale with the Signal-to-Noise Ratio
(SNR)~\cite{Jindal2006}. Conventional channel feedback techniques
utilize pre-defined codebooks to directly quantize and feedback the
channel vector \cite{Love2003,xia2006design,Roh2006,RHS2007}. For a desired accuracy,
codebook size increases exponentially with the number of antennas,
which prevents their application in massive MIMO networks,
especially in frequency selective channels, where feedback is needed
for each coherence bandwidth.
To this end, the complexity and dimensionality
for mMIMO CSI feedback needs to be reduced. 

In~\cite{Love2003} it became clear that when quantizing a complex
vector in $N_t$ dimensions for feedback, one deals with {\it
Grassmannian} quantization---the overall phase of the vector is
irrelevant, and the proper concept of distance is given by the
chordal distance. Vector quantization codebooks can be created, e.g., 
using computer search~\cite{Love2003,xia2006design,Roh2006}.

When the CSI to be quantized is not independent and identically distributed (i.i.d.), and the covariance
matrix is known, one may perform vector quantization conditioned
on the source distribution. A more practical approach, applicable
to any channel covariance, was proposed
in~\cite{love2006limited,xia2006design}: an i.i.d. vector
quantization codebook is constructed, and the channel
covariance structure is imposed on the codebook by deforming it
with the square root of the covariance.

The complexity of designing and quantizing high-dimensional CSI
vectors has been addressed using product
codebooks~\cite{Cheng2010,Yuan2012} and trellis coded line
packing~\cite{au2011trellis,Choi2015}. In these, the $N_t$
dimensions of the vector to be quantized are divided in parts of
a lower dimension $L$, and an $L$-dimensional codebook is used to
quantize the parts. These should be designed based on an
Euclidean distance criterion---when combining the partial
codewords, the phase becomes relevant~\cite{au2011trellis}, and
has to be properly taken into account when combining multiple
low-dimensional codewords to a higher dimensional, either when
computing trellis metrics~\cite{au2011trellis,Choi2015} or
by explicit quantization of combining phases~\cite{Yuan2012}, or by joint designs resulting in codebooks that have good distance properties both from the Euclidean and chordal distance perspective~\cite{Pitaval2014}.

Recently, deep neural networks (DNN) have been applied for mMIMO
CSI feedback~\cite{Gunduz2019, Mashhadi2021, Wang2019}. Compared to
vector quantization
which imposes an overhead that grows linearly with system dimensions,
and also depends on channel sparsity,
DNN-based CSI feedback has promise to overcome complexity, latency and
limited accuracy drawbacks in capturing CSI.

Whether basing feedback on
domain knowledge or DNN,
fundamental theoretical understanding on the feedback framework,
quantization objectives and distortion analysis is needed as to
understand the fundamental limits governing quantized CSI feedback.
This motivates our work.
  
Of particular relevance to our work are \cite{Caire2013, Nam2014,
  Chen2014,Ghosh2012,Liu2017,3GPP,Onggosanusi2018,Song2018,Schwarz19} where modular/cascaded/two-tier/two-stage CSI feedback 
and/or covariance eigenspace quantization is
considered.
The low rank of mMIMO channel covariance matrices was utilized to
reduce effective channel dimensionality.
Modular multiuser-MIMO feedback quantization was suggested in \cite{Caire2013,Nam2014}.
For correlated single user channels $\vech\in\CC^{\Nt}$, the majority of
channel energy is in a low-dimensional subspace
and it is sufficient to feed back signal coordinates in this subspace. Users are clustered such that
users in a cluster are assumed to share a $K$-dimensional subspace,
described by an $N_t\times K$ unitary matrix $\maU_K$.
Instantaneous CSI is then fed back by the users in terms of a
$K$-dimensional \emph{effective channel} vector
$\vecc=\maU_K^\rmH\, \vech$.
In~\cite{Chen2014}, a similar idea of two-tier precoding was pursued.

In~\cite{Liu2017,3GPP,Onggosanusi2018,Song2018}, frequency selective mMIMO channels are considered. Covariance matrices are estimated over the whole bandwidth, and narrowband effective channel CSI is created for multiple subbands. Both wideband covariance
and subband CSI are quantized and fed back to the BS. In~\cite{3GPP,Onggosanusi2018,Song2018}, {\it array architecture aware precoding} is performed, where the feedback is optimized for known dimensions of a planar array. In~\cite{Schwarz19}, two-tier precoding is performed in a multiuser MIMO setting, where one tier is a subspace for a user group, and the other for a user.

To use the modular approach
in~\cite{Caire2013,Nam2014,Chen2014,Liu2017,3GPP,Onggosanusi2018,Song2018}, the
basis matrix ${\bf U}_K$ has to be quantized using a codebook
$\calW$. Covariance matrix and covariance eigenspace
quantization has been considered in
\cite{Ghosh2012,Murga2012,Nam2014,Liu2017,3GPP,Onggosanusi2018,Schwarz2021}.
Matrix codebooks of
unitary matrices characterizing covariance eigenspaces are considered in~\cite{Ghosh2012}. In \cite{Murga2012}, channel Gram matrices are constructed, and information about them is fed back using scalar quantization of matrix elements. 
Hierarchical quantization of $N_t \times K$ unitary (Stiefel) matrices as well as $K$-dimensional subspaces was considered in~\cite{Schwarz20,Schwarz2021,Schwarz22}, where the dimensionality of the subspace {\it decreases} when proceeding in the hierarchy.   
In contrast,~\cite{Liu2017,3GPP,Onggosanusi2018} consider independent
vector quantization of the columns of $\maU_K$, i.e. codebooks of the
form $\calW = \calV^K$, where
$\calV$ is a codebook of $N_t\times 1$ vectors. For
a given quantization accuracy, the size of a vector codebook is
the $K$th root of the size of a matrix codebook, which
significantly reduces quantization complexity. However, when a vector
codebook is used, orthogonality of the matrices cannot be guaranteed.
In the matrix codebook approach of~\cite{Ghosh2012}, orthogonality of
the $K$ columns in $\maU_K$ is always guaranteed. In~\cite{Liu2017},
orthogonality is ensured by sequence design, limiting the vector
codebook size to $N_t$.

To have high descriptive power with manageable complexity,
high-resolution FDD feedback in 5G NR relies on overcomplete vector
quantization codebooks~\cite{3GPP,Onggosanusi2018}. This enables high
precision in  describing
$\maU_K$, at the cost of a potential loss of orthogonality.
With a non-orthogonal basis, good subband effective channel
quantization might not result in good overall quantization.

In this paper, we develop a modular CSI quantization framework for massive MIMO
networks with high precision and low complexity. Design objectives for
both wideband quantization and subband quantization are studied, as
well as codebook construction.
Our major contributions are:
\begin{itemize}

\item We show that the overall quantization distortion
  decomposes into two independent parts. One describes the error of
  the wideband quantization of $\maU_K$, the other captures
  effective subband channel distortion. We find a new criterion
  for measuring $\maU_K$ quantization quality,
  which highlights the importance of improving wideband
  quantization.

\item
  We propose orthonormalized wideband
  precoding (OWP) which implicitly feeds back unitary matrices
  despite using high precision vector codebooks
  for independent quantization of columns
  of $\maU_K$. Thus despite using  
  vector codebooks,
  we guarantee orthogonality of the basis in the fed back matrices, 
  removing the drawback of using a vector codebook.
  Furthermore, we mitigate the loss from vector quantization by 
  proposing a sequential wideband precoding (SWP) scheme.

\item We perform simulations in an FDD mMIMO scenario,
  corroborating the theoretical
  results.
  Following the principles of this paper leads to considerable gains.  
\end{itemize}

This paper is organized as follows. Section \ref{sec:sys} introduces
the system model. Section \ref{sec:ZF} derives the single user
quantization problem from multiuser precoding. Section \ref{sec:msuq}
presents the modular quantization method and discusses
quantization distortion partitioning. Section
\ref{sec:wq} introduces  the wideband quantization schemes OWP and
SWP.
Section \ref{sec:sq} discusses subband effective channel quantization. 
Section \ref{sec:sim} presents
simulation results,
Section \ref{sec:con} concludes the paper.

Notation: $\maX^\rmT$ and $\maX^\rmH$ denote the transpose and the conjugate
transpose of matrix $\maX$, respectively. The Euclidean norm of a vector
$\vecx$ is $\|\vecx\|$. $\Tr\, (\maX)$ denotes the trace of matrix
$\maX$.

\section{System Model}\mabel{sec:sys}

\subsection{Channel Model}

We consider a Frequency Division Duplexing (FDD) multiuser
multicarrier downlink massive MIMO channel in which the transmitter
has $N_t$ antennas, and the $U$ receivers have a single antenna. The
channels are frequency selective. We assume OFDM, and model
frequency selectivity on a subband basis, such that there are $S$
subbands for which CSI is gathered. For simplicity we assume that
the channel $\vech_{u,s} \in \mathbb{C}^{N_t\times1}$ of user $u$ on
subband $s$ is the same for all subcarriers in a subband. 
The received signals of all the users on a subcarrier $c$ belonging
to subband $s$ are
\begin{equation}
\vecy_c=\matH_s \matZ_s\, \vecx_c+ \vecn_c\,. 
\label{eq:MUMIMO}
\end{equation}
Here $\matH_s \in \mathbb{C}^{U  \times N_t}$ is a matrix where the
subband channels of the users are collected s.t. row $s$ is
$\vech_{u,s}^\rmT$.  $\matZ_s\in \mathbb{C}^{N_t
  \times U}$ is a Zero-Forcing (ZF) multiuser beamformer, $\vecx_c \in
\mathbb{C}^{U \times1}$ is the vector of the symbols transmitted to
the users, and ${\bf n}_c$ is independent white Gaussian noise with
variance $N_0$. Transmissions to different users are independent; we
assume $E\{\vecx_c \vecx_c^\rmH \} = \maI_U$. With total transmit
power $P$, the power constraint
then becomes $\Tr\, (\matZ_s\,\matZ_s^\rmH) \le P$.

\subsection{Wideband and Effective Channels}

In FDD, selecting the precoder $\matZ$ requires feeding back
information from the receivers to the transmitter. We assume codebook
based precoding, where each user $u$ independently quantizes  its 
  aggregate CSI $\lbrace \vech_{u,s}\rbrace_{s=1}^S$,
  using a common
  codebook, and then feeds it back to the transmitter.
    As feedback is independent, when discussing
channel feedback, we  suppress the index $u$.
 We shall exploit a division to wideband and subband channels for
efficient feedback.

We assume that the channel of an individual user across subbands is block Rayleigh-faded, so
that it is distributed as a complex Gaussian, with a joint
distribution arising from the multipath structure of the environment.
The user channels on subbands are samples from this distribution, and
a user has perfect knowledge of its own channels. From these, a user
can construct a sample covariance matrix
which is an estimate of
$E_{\vech}\left\{\, \vech\,
\vech^\rmH\right\}$. 
This form of the covariance is used in the literature for capturing the wideband characteristic of the channel \cite{Liu2017, Ghosh2012,Onggosanusi2018}.
However, in this paper  we shall prove that the  normalized
sample covariance
\begin{equation}
  \widetilde\maR = \sum_s \vech_s \,\vech_s^{\rmH}/\left|\vech_s\right|^2
  = \sum_s \tilde\vech_s \, \tilde\vech_s^{\rmH}
  \mabel{eq:maR}
\end{equation}
should be applied for feedback optimization.
  Here we defined the {\it normalized subband channel}
  $\tilde\vech_s=\vech_s/\left|\vech_s\right|$. Note that the sample covariance is user specific.

The aggregate CSI is captured by the sample covariance 
$\widetilde\maR$ describing the \textit{wideband characteristics} of
the channel, and subband specific coordinates describing the subband
channels in the basis given by $\widetilde\maR$. Note that the
normalization in (\ref{eq:maR}) does not change the fact that
eigenvectors of $\widetilde\maR$ span a subspace in which all $h_s$
lie.
We have the eigenvalue decomposition
\begin{equation}\mabel{eq:svd}
\widetilde\maR =\maU \maLambda \maU^\rmH,
\end{equation}
where for ${\rm rank}(\widetilde\maR) = r\leq \Nt$, $\maLambda$ is an $r\times
r$ diagonal matrix of $r$ positive eigenvalues
arranged in descending order, and $\maU=[\vecu_1, \vecu_2, \dots,
  \vecu_r]$ is the tall unitary $N_t\times r$ matrix consisting of the
corresponding eigenvectors $\vecu_j$. We also define
the singular values $\sigma_j = \sqrt{\lambda_j},\, j=1,\ldots,r$, and
the matrix $\maSigma=\diag{\{\sigma_j\}} =
\sqrt{\maLambda}$ of  singular values of
$\sqrt{\widetilde\maR}$.
We have $\maU^\rmH\maU = \maI_r$, the $r$-dimensional
identity matrix, while  $\maU\maU^\rmH = \maPi_r$ is a projector to
the $r$-dimensional eigenspace of $\widetilde\maR$. By construction we have
$\maPi_r \vech_s = \vech_s$. 

Given $\widetilde\maR$, the subband channel vectors can be expressed in two ways
as \textit{ideal subband effective channels}: 
$
\vech_s=\maU\, \vecb_s = \maU\,\maSigma\,\vecc_s
$,
where $\vecb_s$ and $\vecc_s$ are $r \times 1$ vectors, given by 
\begin{eqnarray} 
  \vecb_s&=&\maU^\rmH \vech_s \mabel{eq:bss}  \\
  \vecc_s&=&{\maSigma}^{-1} \vecb_s = {\maSigma}^{-1} \maU^\rmH \vech_s\,.
 \mabel{eq:css}
\end{eqnarray}
We have $\|\vecb_s\| = \|\vech_s\|$, while the transformation to $\vecc_s$
does not preserve norm, $
\vecc_s^\rmH \maLambda \vecc_s = \|\vech_s\|^2
$.

The sample covariances of $\vecb$ and $\vecc$ with the same
normalization as in $\widetilde\maR$ are
\be
\sum_{s=1}^S  \vecb_s\vecb_s^\rmH = \maLambda\,,~~~~
\sum_{s=1}^S  \vecc_s\vecc_s^\rmH = \maI\,.
\mabel{eq:iid}
\ee
The coordinates in $\vecb_s$ are thus
{\it  independently but non-identically
  distributed (i.n.i.d.)} while the effective channel coordinates $\vecc_s$ are i.i.d.

\section{Zero Forcing Precoding with Incomplete CSI at Tx}\mabel{sec:ZF}

The transmitter constructs a multiuser ZF precoder per
subband $s$, used on each subcarrier channel (\ref{eq:MUMIMO})
belonging to the subband, based on {\it incomplete
  CSI} fed back by the users to the Tx. We divide the CSI to Channel
Direction Information (CDI), 
quantizations $\hat\vech_{u,s}$ of the
normalized channel vectors $\tilde\vech_{u,s}$, and Channel Amplitude
Information (CAI) $\hat q_{u,s0} > 0$, quantizing $|\vech_{u,s}|$.
The CDI takes values in a user-specific {\it quantization codebook},
$\hat\vech_{u,s}\in \calH$.

\subsection{Bound on Expected Rate}

The CDI of the users on the subband are collected to the $U \times
N_t$ matrix $\hat\maH_s$, where row $u$ is $\vech_{u,s}^\rmT$.
We also divide the precoder to a normalized part $\widetilde\matZ$ and
a diagonal power allocation part $\matA_s$ with diagonal elements
$a_{u,s}$, as
\be
 \matZ_s = \widetilde\matZ_s \matA_s\,, 
 \ee
were $\left|\vecz_{u,s}\right|=1$. 
The power constraint now reads
$\Tr\left({\widetilde\matZ_s}^\rmH\widetilde\matZ_s \, \matA_s^2\right) \leq
P$.
As the subspace spanned by a set vectors equals the subspace spanned
by the corresponding normalized versions, $\widetilde\matZ$ does not
depend on CAI, we have $\widetilde\matZ_s =
\widetilde\matZ_s\left(\hat\maH_s\right)$.
The ZF precoding vector $\tilde\vecz_u$ is defined as a unit
norm vector in the subspace spanned by
$\hat\maH_s^*$, s.t. $\vech_{v,s}^\rmT \vecz_u=0$ for all $v\neq u$.

Assuming that $\vecx_c$ in (\ref{eq:MUMIMO}) are independent, and
independent from the noise, the Signal-to-Interference-plus-Noise
Ratio (SINR) on subband $s$ at user $u$ becomes:
\columnversion{\be
\gamma_{u,s}  = \frac{\left|\vech_{u,s}^\rmT \vecz_u\right|^2}{
  \sum_{v\neq u} \left|\vech_{u,s}^\rmT \vecz_v\right|^2  + N_0}
= \frac{\rho_{u,s}\, \left|\tilde\vech_{u,s}^\rmT \tilde\vecz_u\right|^2}{
  1 + \rho_{u,s} \sum_{v\neq u} \frac{a_v^2}{a_u^2} \left|\tilde\vech_{u,s}^\rmT \tilde\vecz_v\right|^2 }~,
\ee}{\begin{eqnarray}
\gamma_{u,s}  &=& \frac{\left|\vech_{u,s}^\rmT \vecz_u\right|^2}{
	\sum_{v\neq u} \left|\vech_{u,s}^\rmT \vecz_v\right|^2  + N_0}\cr
&=& \frac{\rho_{u,s}\, \left|\tilde\vech_{u,s}^\rmT \tilde\vecz_u\right|^2}{
	1 + \rho_{u,s} \sum_{v\neq u} \frac{a_v^2}{a_u^2} \left|\tilde\vech_{u,s}^\rmT \tilde\vecz_v\right|^2 }~,
\end{eqnarray}
}
where
$\rho_{u,s} =
a_u^2\,\left|\vech_{s,u}\right|/N_0$ is the Signal-to-Noise Ratio
(SNR) of user $u$ on subband $s$. Assuming Gaussian codebooks
the rate of user $u$ on subband $s$ is lower bounded as~\cite{Jindal2006}
\columnversion{\be
 R_{u,s} = \ln\left(1+ \gamma_{u,s}\right) \geq 
\ln\left(1+\rho_{u,s}\, \left|\tilde\vech_{u,s}^\rmT
\tilde\vecz_u\right|^2\right) -
\ln\left(1 + \rho_{u,s} \sum_{v\neq u} \frac{a_v^2}{a_u^2}
\left|\tilde\vech_{u,s}^\rmT \tilde\vecz_v\right|^2\right)\,.
\label{eq:Rbound}
\ee}{\begin{eqnarray}
 R_{u,s} &=& \ln\left(1+ \gamma_{u,s}\right) \geq 
\ln\left(1+\rho_{u,s}\, \left|\tilde\vech_{u,s}^\rmT
\tilde\vecz_u\right|^2\right) \cr &&
-\ln\left(1 + \rho_{u,s} \sum_{v\neq u} \frac{a_v^2}{a_u^2}
\left|\tilde\vech_{u,s}^\rmT \tilde\vecz_v\right|^2\right)\,.
\label{eq:Rbound}
\end{eqnarray}
}
We assume that a
user grouping principle is applied to select the $U$ users served in the
subband, e.g., users with near-orthogonal channels $\hath_{u,s}$ may
be selected. This results in a distribution of $\beta_u =
\left|\hat\vech_{u,s}^\rmT \tilde\vecz_u\right|^2\in [0,1]$. We furthermore
assume a power allocation principle that jointly selects the
amplitudes $a_{u,s}$ for all the users, based on the CAI $q_{u,s}$ and
the realizations of $\beta_u$.

The true channel direction can expressed in terms of the quantized CDI as
\be
\tilde\vech_{u,s} = \sqrt{1-\epsilon^2}\, \hat\vech_{u,s}
+ \epsilon_{\Vert}\, \vece_{\Vert} + \epsilon_{\perp}\,
\vece_{\perp}\,,
\label{eq:quantiErrors}
\ee
where quantization error has two parts,
one in the subspace spanned by the other user's channels
$\left\{\vech_{v,s}\right\}_{v\neq u}$, another in the subspace orthogonal to all
user's channels. The directions of these are given by the unit vectors
$\vece_{\Vert}$, and
$\vece_{\perp}$, respectively, and the magnitudes add up as
$\epsilon_{\Vert}^2 + \epsilon_{\perp}^2 = \epsilon^2$.  In the wanted
signal component in (\ref{eq:Rbound}) we thus have
 $$
\left|\tilde\vech_{u,s}^\rmT \tilde\vecz_u\right|^2 =
(1-\epsilon^2)\left|\hat\vech_{u,s}^\rmT \tilde\vecz_u\right|^2 =
(1-\epsilon^2)\beta_u\,,
 $$
and in the interference-part 
 $$
\left|\tilde\vech_{u,s}^\rmT
\tilde\vecz_v\right|^2
 = \epsilon_1^2 \left|{\vece_{\Vert}}^\rmT \tilde\vecz_v\right|^2
 = \epsilon_1^2 \delta_{u,v} \leq \epsilon^2 \delta_{u,v}
 \,,$$
 for some $\delta_{u,v}\in[0,1]$.

In~\cite{Jindal2006}, the distribution of quantization error was
derived for the case $U=N_t$, with channels chosen uniformly at
random, equal power allocation and random vector quantization, leading
to a lower bound on the rate. 
This can be extended to any $U$, user
grouping, power allocation and quantization codebook
as follows.

For a user of interest, we average over the possible channel states
$\vech_{s}$, and all possible other user's channels given by a
grouping principle. We assume that the quantization of
$\tilde\vech_{u,s}$ in $\calH$ is {\it homogeneous}; the magnitude
$\epsilon$ of the quantization error is independent of
$\hat\vech_{u,s}$ and
the directions $\vece_{\Vert}$, $\vece_{\perp}$. 
From the variables effecting the SINR, 
the power
allocation amplitudes $a_v$, 
the SNR $\rho$
and the inner product $\beta$ are thus jointly distributed, while
$\epsilon$ and the
$\delta_{u,v}$ are independent.

With the definitions above, the rate bound (\ref{eq:Rbound}) becomes
\columnversion{
\be
 R_{u,s}\geq 
 \ln\left(1+ (1-\epsilon^2)\rho\,\beta\right) -
\ln\left(1 + \epsilon^2 \rho \sum_{v\neq u}
\frac{a_v^2}{a_u^2} \delta_{u,v} \right)
 := R_1 + R_2\,.
\label{eq:Rbound2}
\ee}{$$
R_{u,s}\geq 
\ln\left(1+ (1-\epsilon^2)\rho\,\beta\right) -
\ln\left(1 + \epsilon^2 \rho \sum_{v\neq u}
\frac{a_v^2}{a_u^2} \delta_{u,v} \right)\,.
$$}
where we dropped the subscript from $\rho$ and $\beta$. 
The first term $R_1$ can be further bounded 
as
\columnversion{
\be
R_1 \geq  \ln\left(1+\rho\,\beta\right)  +
\ln\left(1-\epsilon^2\right)
 \,\geq\, \ln\left(1+\rho\,\beta\right)  - \frac{\epsilon^2}{1-\epsilon^2}
 \,\geq\, \ln\left(1+\rho\,\beta\right)  - \epsilon^2\,. 
\ee}{\begin{eqnarray*}
R_1 &\geq&  \ln\left(1+\rho\,\beta\right)  +
\ln\left(1-\epsilon^2\right) \cr
&\geq& 
\ln\left(1+\rho\,\beta\right)  - \frac{\epsilon^2}{1-\epsilon^2} \,\geq\, \ln\left(1+\rho\,\beta\right)  - \epsilon^2\,. 
\end{eqnarray*}}
We used
the inequality $x/(1+x)
\leq \ln (1+x)$, valid for $x\geq -1$~\cite{Landau1934}. 
When averaging over the channels, the second term can be
bounded using Jensen's inequality. Defining $C_1= E_{\beta,\rho}\left\{\ln(1+\rho\,\beta) \right\}$ and 
$C_2= E_{\rho,a,\mathbf{\delta}}\left\{\rho\sum_{v\neq u}
\frac{a_v^2}{a_u^2} \delta_{u,v} \right\}$, 
we get a lower bound for the expected rate:
\be
E_{\beta,\rho,a,\mathbf{\delta},\epsilon}\left\{R\right\}\geq
C_1 - E_{\epsilon}\left\{\epsilon^2\right\} -\ln\left(1 + C_2 \, E_{\epsilon}\left\{\epsilon^2\right\} \right)\,.
\label{eq:RexpectBound}
\ee

\subsection{Quantization Distortion}

The \emph{chordal distance} between two vectors
$\mathbf{x}$ and $\mathbf{y}$ is 
\begin{equation}\mabel{eq:chordal}
d(\mathbf{x},\mathbf{y})=\sqrt{1-\frac{|\mathbf{x}^\rmH\mathbf{y}|^2}{\|\vecx\|^2 \|\mathbf{y}\|^2}}\,.
\end{equation}

  According to (\ref{eq:quantiErrors}), the quantization error
is given by the chordal distance, $\epsilon^2 =
d^2(\tilde\vech_{s},\hat\vech) = d^2(\vech_{s},\hat\vech)$.
From (\ref{eq:RexpectBound}) it follows that the lower bound on rate
is maximized if the codebook $\calH$ for quantizing $\vech_s$ is
chosen to minimize the {\it quantization distortion}
\begin{equation}\mabel{eq:goal}
D_\calH = E_{\vech_s} \left\{\underset{\hath\in \calH}{\mathrm{min}}\,
d^2(\vech_s,\hath)\right\}
= \sum_{\hath\in\calH} E_{\Voronoi(\hath)} \left\{d^2(\vech_s,\hath)\right\}
\,.
\end{equation}
In the second expression, the distortion has been written as a sum
over the division of the domain of $\vech_s$ to Voronoi cells
$\Voronoi(\hath) = \{ \vech \mid \arg\min_{\vecq \in\calH}\,
d(\vech,\vecq) = \hath \}$ centered at the codewords  $\hath\in\calH$. 

Our goal is to develop quantization schemes that minimize distortion.
Note that the distortion expressly considers quantization of
  channel direction only. For the bound (\ref{eq:RexpectBound}), the quantization error in $|\vech_s|$ irrelevant. The decoupling of CDI and CAI quantization is a consequence of ZF.

\section{Modular Single User CSI Quantization}\mabel{sec:msuq} 

In this section, the overall process of the developed modular CSI
quantization approach for a single-user codebook $\calH$ will be
presented, followed by a discussion on partitioning quantization
distortion between wideband and subband quantization.

\subsection{Basic idea and earlier works of modular quantization}

If the subband channels $\mathbf{h}_s$ come from an arbitrary
distribution,
their quantization and feedback can be highly complex. However, if we
can assume that the channels from different antennas are correlated,
it is possible to apply modular quantization~\cite{Caire2013,Nam2014,Chen2014,Liu2017,Song2018} and reduce complexity considerably. 
In this section we shortly review some of the known results and then present our approach.

\subsubsection{Modular quantization with ideal wideband quantization}
Let us assume
that the covariance matrix $\mathbf{R}$ is fixed and we have a budget
of bits for feeding back wideband statistical data of the channel
pertaining to $\mathbf{R}$, and a separate budget for feeding back
subband specific CSI pertaining to the coordinates $\mathbf{c_s}$.
In the extreme case with unlimited wideband feedback, the BS knows
$\maU$ and $\mathbf{\Lambda}^{1/2}$ perfectly. For limited feedback, we make the {\it rank-$K$ approximation} of the covariance matrix. 
We denote with $\maU_K$ the $N_t\times K$ matrix consisting of columns of $\maU$ ordered by the size of the corresponding eigenvalues, and 
$\maLambda$ and $\maSigma$ are now $K\times K$ diagonal matrices with the $K$ largest eigen- and singular values. 

Ideal subband quantization would then proceed as follows. Given $\maU_K$, the \emph{subband effective channel} vectors can be expressed in two ways as
\columnversion{
\begin{equation}
    \vecb_s =\maU_K^\rmH \vech_s\,,~~~~
  \vecc_s={\maSigma}^{-1} \vecb_s = {\maSigma}^{-1} \maU_K^\rmH \vech_s\,,
   \mabel{eq:deconstruct}
   \end{equation}
   }{
\begin{eqnarray} 
  \vecb_s&=&\maU_K^\rmH \vech_s \mabel{eq:bsss}  \\
  \vecc_s&=&{\maSigma}^{-1} \vecb_s = {\maSigma}^{-1} \maU_K^\rmH \vech_s\,,
   \mabel{eq:deconstruct}
 \end{eqnarray}
 }
  where $\vecb_s$ and $\vecc_s$ are length 
$K$ random vectors. Note that $\vecc_s$ is typically not unit norm. However, as the bound (\ref{eq:RexpectBound}) is in terms of chordal distance and does not depend on $|\vech_s|$, we may normalize $\vecc_s$. 
The user then quantizes $\vecc_s$ to $\hatc_s$ using a $K$-dimensional Grassmannian line codebook $\hat{\mathcal{C}}\subset\calG_{\CC}(K,1)$
and feeds this information to the BS.
The BS can construct an estimate of $\vech_s$ as
\begin{equation}\label{eq:reconstruct}
 \hath_s=\maU_K\maSigma\hatc_s.
\end{equation}
The key point is that the modular approach reduces the dimension of subband quantization from $N_t$ to $K$. 
Due to the unitarity of $\maU_K$, good quantization on the effective channel should lead in good quantization on the actual channel, as $\maU_K$ captures most of the energy of $\vech_s$. 

\subsubsection{Wideband quantization}

For wideband feedback, either a codebook of orthogonal matrices~\cite{Ghosh2012,Schwarz19} or a
vector codebook $\hat{\mathcal{C}}_{\vew}$ for quantizing the
$N_t\times 1$ columns in $\maU_K$  \cite{Song2018,Onggosanusi2018} can be used. A scalar codebook is used
for quantizing the elements in $\maSigma$ \cite{Onggosanusi2018}.
After wideband CSI feedback,
the BS has a matrix $\maV$, which is a quantized version of $\maU_K$, and
$\hatSigma$, which is a quantized version of $\maSigma$.
Subband quantization now proceeds as in
(\ref{eq:deconstruct},\ref{eq:reconstruct}) but replacing $\maU_K$ with $\maV$ and $\maSigma$ with $\hatSigma$.
However, if the vector codebook
$\hat{\mathcal{C}}_{\vew}$ is overcomplete, as for example in the high
resolution alternative of 5G NR~\cite{3GPP}, there is a high
probability that $\maV$ is not unitary, and the connection between the channel and effective channel quantization is partially broken. In the following we will show how we can avoid this problem.

\subsection{Outline of the Suggested CSI Quantization Method}

In this section we will shortly describe 
our approach to the quantization problem. In later sections we will elaborate on each part of the quantization process.

While conventionally wideband feedback is based on quantizing the
covariance matrix $\maR=E_{\vech}\left\{\vech
  \vech^{\rmH}\right\}$~\cite{Liu2017}, or a normalized version
  $E_{\vech}\left\{\vech
  \vech^{\rmH}\right\}/E_{\vech}\left\{\|\vech\|^2\right\}$~\cite{Onggosanusi2018},
      we instead quantize $\widetilde\maR$ of (\ref{eq:maR})
and its singular value decomposition. The motivation for this will be given in Section \ref{ss:distortion}.

Aiming at  $K$-dimensional effective channel feedback, 
we proceed by
quantizing $\maU_K$ column by column using some vector quantization
codebook $\hat{\mathcal{C}}_{\vew}$. As a result the user has now an
$N_t\times K$ matrix $\maV$ consisting of quantized norm $1$ vectors,
which are fed back to the BS. 
These vectors are not necessarily orthonormal.

However, we assume that the users and the base-station {\it have
  agreed on a method to orthonormalize} the vectors $\maV$
to a set of column vectors $\maW$ with $\maW^\rmH\maW = \maI_K$ and ${\rm span}(\maW) = {\rm span}(\maV) $. 
This method is independent of the structure of
$\maV$ and is assumed to be shared at the same time as the codebook
$\hat{\mathcal{C}}_{\vew}$. Both the user and the BS perform this operation, after which they both have the same matrix $\maW$.
Exactly the same number of bits
can be used to feed back this matrix as feeding back the original
$\maV$. The matrix $\maW$ is now used in place of $\maV$ for the rest
of the quantization.

As $\maV$ is transformed to $\maW$  
the singular values computed from (\ref{eq:svd}) do not characterize
the relative weight of the columns in $\maW$ when describing
subbands. Instead, for each column $\mathbf{w}_j$ in $\maW$,
the user calculates {\it wideband amplitudes}
\begin{equation}\mabel{eq:wbamp}
\sigma_j=\sqrt{\mathbf{w}_j^{\rm H} \widetilde\maR \mathbf{w}_j}
\end{equation}
and feeds back their scalar quantized versions $\hat{\sigma}_j$
constituting the diagonal matrix $\hat{\maSigma}$.

After this, BS and user share the matrix $\maW$ and the quantized wideband amplitude matrix $\hat{\maSigma}$. Using these matrices, the
subband feedback can be performed as in Equations
\eqref{eq:deconstruct} and \eqref{eq:reconstruct}, replacing $\maU_K$
and $\maSigma$ with $\maW$ and $\hat{\maSigma}$.
The reconstruction of the channel
can be done as
$$
\hath_s=\maW\hat{\maSigma}\hatc_s.
$$

When $\maW$ and $\hat{\maSigma}$ are fixed, then any quantization codebook $\mathcal{C}\subset \mathbb{C}^K$, designed to quantize a $K$-dimensional effective channel $\vecc_s=\hat{\maSigma}^{-1}\maW^{\rmH}\mathbf{h_s}$, induces a codebook~\cite{love2006limited,xia2006design}
\begin{equation}
    \calB= \left\{ \frac{\hat{\maSigma}\hatc}{|\hat{\maSigma}\hatc| } \,\,\Big|\,\, \hatc \in \calC \right\} \,,
    \label{eq:calB}
\end{equation}
which quantizes the $K$-dimensional effective channel $\maW^H\vech_s=\vecb_s.$
Furthermore it also induces a codebook for the actual $N_t\times 1$ channel $\vech_s$ by
\begin{equation}
  \calH =  \maW\,  {\calB} =  \{\hath = \maW\,\hatb\, \mid\,\hatb \in {\calB} \}\,.
\mabel{eq:calH}
\end{equation}

The whole quantization process  can now be divided to  three steps:
spatial compression and wideband quantization, dimensionally reduced
subband quantization, and reconstruction. The users and the BS share information about the quantization process, and the codebooks $\cal H, B, C$. A flowchart is presented in Fig.~\ref{fig:csi}. The flowchart also contains terms that are not yet defined, but we will elaborate on them later on.

\begin{figure}[!t]
\centering
\includegraphics[width=8cm]{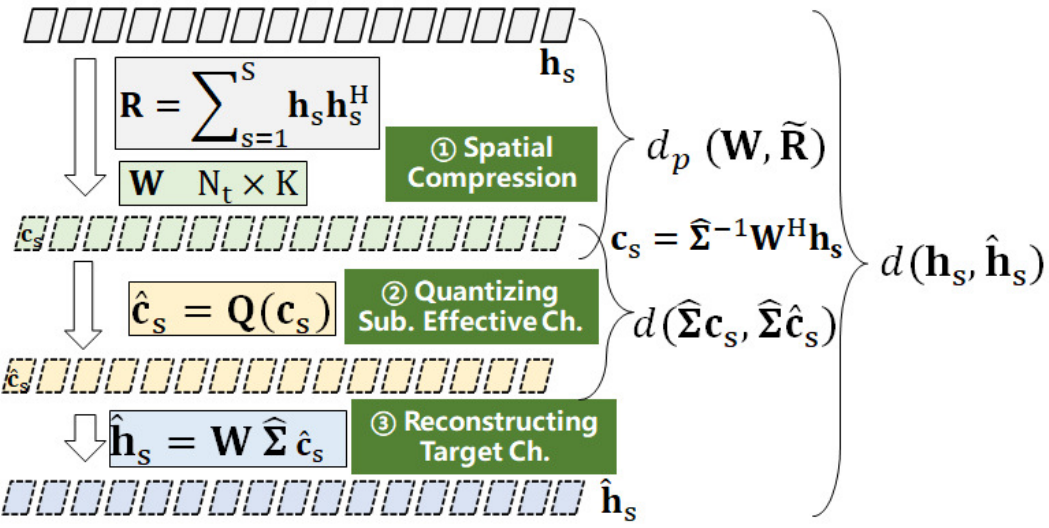}
\caption{Flowchart for general process of CSI quantization.}\mabel{fig:csi}  
\end{figure}

\subsection{Quantization Distortion Partitioning}\mabel{ss:distortion}	
To design a modular quantization scheme, we should understand how the
wideband and subband components in the quantization procedure affect
the overall quantization distortion.
We start by analyzing the overall subband channel distortion 
when using a generic $\Nt \times
K$ wideband precoder $\maV$, resulting from a wideband quantization procedure.

As orthogonality of vectors may be lost in
quantization, a generic quantized $N_t \times K$ wideband beamforming matrix $\maV$ may not be
orthogonal. We can orthogonalize it as: 
\begin{equation}\mabel{eq:wwo}
\maV = \maW \maC^{-1}
\end{equation}
where $\maW$ is an $\Nt\times K$ Stiefel matrix
satisfying $\maW^\rmH \maW=\maI$ and
$\maC$ is a $K\times K$ matrix fulfilling $ \maC\maC^\rmH =
\left(\maV^\rmH\maV \right)^{-1} $. If Gram-Schmidt orthogonalization
is used, (\ref{eq:wwo}) is the QR-decomposition of $\maV$, with $\maC$
the upper triangular Cholesky factor of the basis correlation matrix.
The wideband beamformer defines a projection operator
\be
\maPi_W =  \maV \left(\maV^\rmH\maV \right)^{-1} \maV^\rmH  = \maW\maW^\rmH 
\mabel{eq:PiW}
\ee
mapping $\mathbb{C}^{N_t}$ to the $K$-dimensional
subspace spanned by the columns of $\maV$ (and $\maW$).

We have a codebook $\calH$ to quantize $N_t$-dimensional subband channels $\vech :=\vech_s$. 
According to \mbox{\eqref{eq:chordal}-\eqref{eq:goal}} we are
interested in how well the elements of $\calH$ quantize $\vech$ in
terms of chordal distance. Hence, all the codewords $\hat{\vech}\in
{\calH}$ satisfy $\|\hat{\vech}\| = 1$.

We can now decompose $\vech$ to a component $\vech_{\Vert}$ lying in the column space
of $\maW$, and to a component $\vech_{\perp}$ in the perpendicular subspace, lost in wideband quantization;
\be
\vech = \maPi_W \vech + (\maI - \maPi_W)\vech := \vech_{\Vert}+ \vech_{\perp}~.
\ee
We use shorthand notations
$\tilh = \vech/\|\vech\|$,
$\tilh_{\Vert} =\maPi_\maW \tilh = \vech_{\Vert}/\|\vech\|$ and
$\tilh_{\perp} = \vech_{\perp}/\|\vech\|$. These
fulfill
\begin{equation}\mabel{is1}
  \|\tilh\|^2 = \|\tilh_{\Vert}\|^2+ \|\tilh _{\perp}\|^2=1.
\end{equation}
As $\hath$ is in the span of $\maW$ we have $\tilh^\rmH
\hath= \tilh_{\Vert}^\rmH \hath$.  The
distortion can thus be decomposed as
\begin{equation}
  d^2(\vech,\hat\vech) =
  \|\tilh_{\Vert}\|^2~d^2(\vech_\Vert,\hat\vech)
+\|\tilh _{\perp}\|^2\,.
\mabel{lemma:comp}
\end{equation}
There is a component related to the distortion in the subspace spanned by $\maV$, and a component arising from the part lost in wideband quantization.

For a $\maW$ and a normalized covariance $\maR$ with $\Tr\,(\maR) =
1$, we define the \emph{projection distortion}
\be
 d_p(\maW,\maR) = 1-\Tr\left(\maPi_\maW\, \maR \right)\,.
\mabel{eq:projdist}
\ee
With $\maU$ and $\maSigma$ the eigenvector and singular value matrices
of $\maR$, we see that this is nothing but the squared Euclidean distance
between $\maU\maSigma$ and its projection to $\maW$:
\be
d_p\left(\maW,\maU\maSigma^2\maU^\rmH\right)
= \left\Vert \maU\maSigma - \maPi_\maW\maU\maSigma \right\Vert_F^2\,.
\ee
For $\maW$ of (\ref{eq:wwo}) an orthogonalization of $\maV$, we have $d_p(\maW,\maR) = d_p(\maV,\maR)$.

As the projection operator $\maPi_W$ maps 
$\mathbb{C}^{N_t}$ to the $K$-dimensional subspace spanned by the columns of $\maW$, the projection distortion measures the amount of power of $\maR$ outside the column space of the wideband beamformer $\maW$. To minimize distortion, the wideband beamformer $\maV$ should capture as much power as possible from the covariance matrix $\maR$ in its column space. 
As such, the projection distortion measures the accuracy of wideband quantization. 

We can now formulate the central result of this paper, governing the
partition of quantization distortion between wideband and subband
quantization:

\begin{proposition}\mabel{prop:overall}
Given an ensemble of random $\Nt\times 1$ vectors $\vech$, an $\Nt\times K$ matrix
$\maW$ and a quantization codebook ${\calH}$ in the subspace spanned
by $\maW$, the quantization distortion is
$$
D_\calH  = 
E_{\vech}\left\{ \|\tilh_{\Vert}\|^2\,
  \underset{\hath\in\calH}{\min}~d^2\!\!\left(\vech_{\Vert},\hath\right)
  \right\}+ d_p(\maW,\widetilde\maR)
$$
where $\widetilde\maR$ is the covariance (\ref{eq:maR}) of {\it
  normalized} vectors $\tilh$, $d_p$ is the projection distortion of
(\ref{eq:projdist}), and $\vech_\Vert= \maPi_W\vech$, with $\maPi_W$ the
projector to the column space of $\maW$. 
\end{proposition}

\begin{IEEEproof}
\columnversion{In}{Following} (\ref{eq:goal}), we first compute the
expectation over $\vech$ in a Voronoi cell $\Voronoi$ with centroid
$\hath$. From (\ref{lemma:comp}) we get \columnversion{ $
  E_{\Voronoi} \left\{  d^2(\vech,\hat\vech) \right\} =
  E_\Voronoi\left\{  \|\tilh_{\Vert}\|^2  d^2(\vech_\Vert,\hat\vech) \right\}
 + E_\Voronoi\left\{\|\tilh _{\perp}\|^2\right\}
$.
}{
\begin{align*}
  E_{\Voronoi} \left\{
  d^2(\vech,\hat\vech)
  \right\} =
  E_\Voronoi\left\{
  \|\tilh_{\Vert}\|^2
  d^2(\vech_\Vert,\hat\vech)
  \right\}
 + E_\Voronoi\left\{\|\tilh _{\perp}\|^2\right\}.
\end{align*}
}
It follows from~\eqref{is1} that
\columnversion{
$
  E_\Voronoi\left[\|\tilh _{\perp}\|^2\right] =
  1 -E_\Voronoi\left[\|\tilh_{\Vert}\|^2\right] =
  1 - E_\Voronoi\left[\Tr(\maPi_W \tilh \tilh^\rmH) \right]
$.
}{
\begin{align*}\nonumber
E_\Voronoi\left[\|\tilh _{\perp}\|^2\right]& = 1
-E_\Voronoi\left[\|\tilh_{\Vert}\|^2\right] =  1 -
E_\Voronoi\left[\Tr(\maPi_W \tilh \tilh^\rmH) \right]\,.
\end{align*}
}
Since the expected value commutes with trace and multiplication with
constant matrices, the expectation acts only on $\tilh \tilh^\rmH$ in
the last term.  
The result follows by summing over all Voronoi cells, and using
(\ref{eq:maR}).
\end{IEEEproof}

Here $\widetilde\maR$ is indeed the covariance of the directions
$\tilh$ of the subband channel vectors, the  channels are normalized {\it before} computing the covariance.
It is normalized as:
\be
\Tr\,(\widetilde\maR)= E_\vech\left\{ \Tr\, (\tilh\tilh^\rmH) \right\}
= E_\vech\left\{  \tilh^\rmH\tilh \right\} = 1\,,
\ee
thus the projection distortion is well-defined for it.

\columnversion{Note}{It is worth noting} that minimum distortion
quantization in Proposition \ref{prop:overall} is performed in the codebook $\calH$ of
$\Nt$-dimensional vectors. To reduce subband quantization complexity, quantization should be performed directly on
$K$-dimensional effective channels $\vecb$ or $\vecc$. We thus use a
codebook $\calB$ to quantize the $K\times 1$ effective channels. With wideband precoder $\maV$, the
induced codebook for $\vech$ is
$
  \calH =  \maV\,  {\calB} =  \{\hath = \maV\,\hatb\, \mid\,\hatb \in {\calB} \}\,.
$
$K$-dimensional minimum chordal distance quantization 
\be
 \hatb = \arg\min_{\hatb\in\calB} \{\,d(\vecb,\hatb)\,\}
\mabel{eq:b-quant}
\ee
may now be performed, given effective channels $\vecb$. 
For the resulting $N_t$-dimensional codeword $\maV\hatb$ to be the same as the minimum distance codeword $\arg\min_{\hath\in\maV\calB}d(\vech,\hath)$ in $N_t$ dimensions, the mapping $\maV$ between $N_t$ and $K$ dimensions has to {\it preserve chordal distance}.
We have:

\begin{lemma}\mabel{lem:isometry}
Consider a linear mapping $\maV\in \CC^{\Nt\times K}$ mapping
$\vecb\in\CC^K$ to $\vech = \maV\vecb\in \CC^{\Nt}$. $\maV$ is an
isometry w.r.t. the chordal distance,
\columnversion{
$d(\vech_1,\vech_2) = d(\vecb_1,\vecb_2)$
}{
\be
 d(\vech_1,\vech_2) = d(\vecb_1,\vecb_2),
 \ee
 }
if and only if $\maV^\rmH\maV \sim \maI_K$. 
\end{lemma}
\begin{IEEEproof}
Let us assume that  $\maV^\rmH\maV \sim \maI_K$.
Recalling (\ref{eq:chordal}) and noticing
that \columnversion{$|(\maV\vech_1)^{\rmH}\maV\vech_2|=|\vech_1^{\rmH}\vech_2|$}{$$|(\maV\vech_1)^{\rmH}\maV\vech_2|=|\vech_1^{\rmH}\vech_2|
$$}
it follows that orthogonality implies isometry. To complete the proof, we need to show that non-orthogonality implies non-isometry. For this, first assume that $\maV^\rmH\maV = \maC$, with an off-diagonal element $c_{ij}\neq 0$ for some $i\neq j$. Take the $K$-dimensional unit vectors $\vece_i$ and $\vece_j$. These are orthogonal\columnversion{;}{, such that} $d(\vece_i,\vece_j)=1$. In the $\Nt$-dimensional space, $\maV\vece_i$ and $\maV\vece_j$ have a non-vanishing inner product. This is not changed by normalization, thus $d(\maV\vece_i,\maV\vece_j)<1$. If $\maC$ is diagonal, but not proportional to identity, there exists some $i,j$ for which $c_{ii}\neq c_{jj}$. \columnversion{The}{Now the} chordal distance between the $K$-dimensional vectors $\vece_i$ and $\left(\vece_i + \vece_j\right)/\sqrt{2}$ is not preserved by $\maV$.    
\end{IEEEproof}

For an orthogonal wideband precoder $\maW$, we define the subband quantization
distortion of (\ref{eq:b-quant}) as 
\be
 D_\calB = E_{\vech}\left\{ \min_{\hatb\in\calB} d^2(\maW^\rmH
 \vech,\hatb) \right\}\,.
 \mabel{eq:DcalB}
\ee
From the results above it now follows

\begin{corollary}\mabel{cor:bounds}
For quantization codebook $\calH = \maW\calB$ with orthogonal
$\maW$, the overall distortion is  upper and lower bounded as:
\begin{align*}
  d_p(\maW,\widetilde\maR) ~\leq~ D_\calH
  ~\leq~
  D_\calB+d_p(\maW,\widetilde\maR)\,.
\end{align*}
\end{corollary}

\begin{IEEEproof}
The first inequality is a direct corollary of Proposition \ref{prop:overall}.
To prove the second inequality, we again use Proposition \ref{prop:overall}. Since $\|\tilh_{\Vert}\|^2 \leq 1$ we have
$$
E_{\vech}\left\{ \|\tilh_{\Vert}\|^2\,
  \min_{\hath\in\calH}~d^2\!\!\left(\vech_{\Vert},\hath\right) \right\}
\leq 
E_{\vech}\left\{\min_{\hath\in\calH}~d^2\!\!\left(\vech_{\Vert},\hath\right) \right\}\,.
$$
The second inequality now follows by using Lemma~\ref{lem:isometry}.
\end{IEEEproof}

\begin{figure}
\centering
\includegraphics[width=80mm]{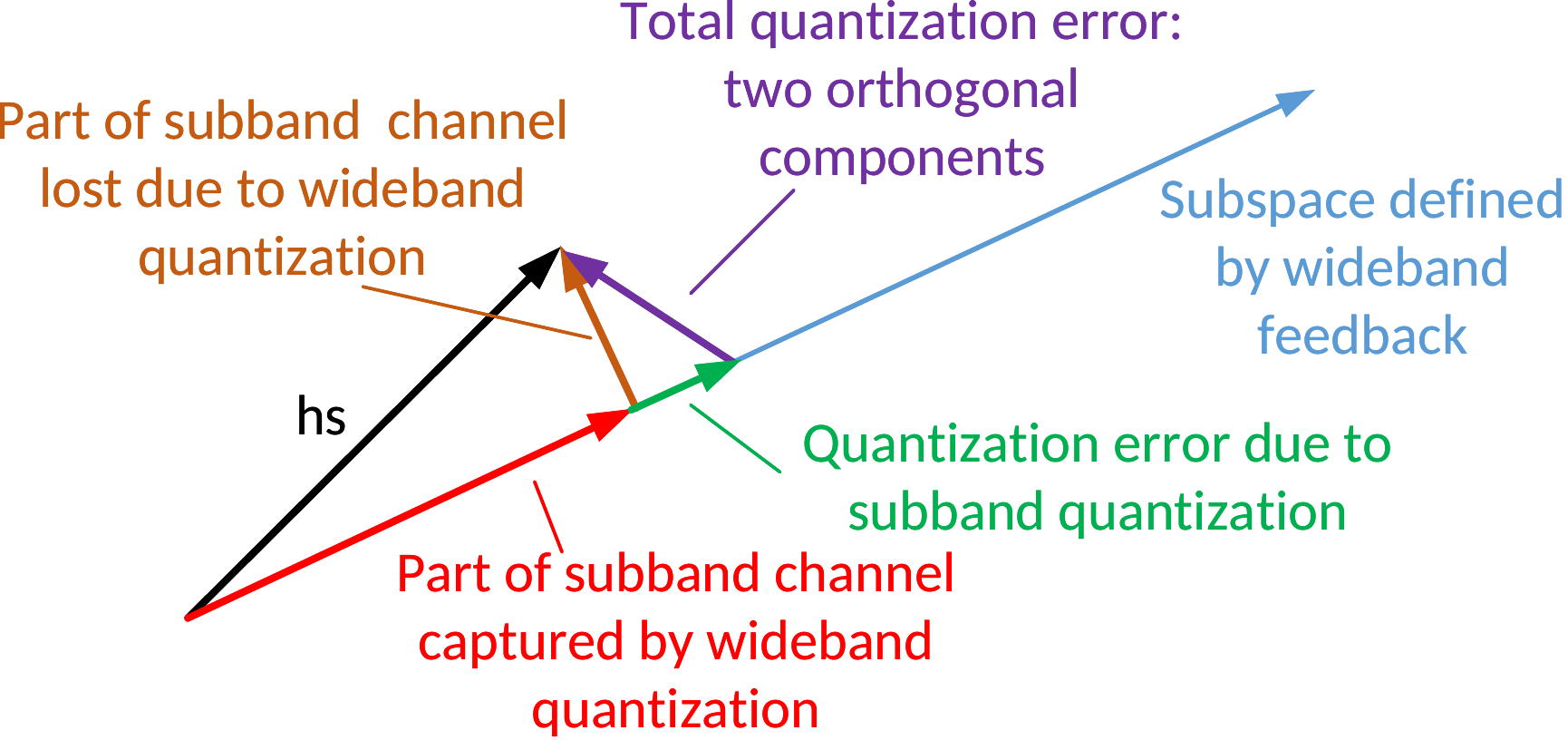}
\caption{Quantization error partition.}
\mabel{fig:2} 
\end{figure}

Fig.~\ref{fig:2}, and the following conclusions summarize our findings
of the theoretical basis for modular CSI quantization:

\begin{itemize}

\item Quantization performance depends on two independent parts:
  wideband distortion, characterized by $d_p(\maW,\widetilde\maR)$,
  and a subband distortion, which depends on the codebook $\calH$.
  Improving either part improves the overall quantization
  (Proposition~\ref{prop:overall}).

\item If $\maW$ is fixed, even perfect subband quantization does not
  provide perfect overall quantization due to $d_p(\maW,\maR)$.
  
\item With orthogonal $\maW$, improving subband effective channel
  quantization by reducing $D_\calB$ directly reduces overall
  distortion (Corollary~\ref{cor:bounds}).
\end{itemize}

Note that the analysis above reveals how the subband
distortion is given by the orthogonal component $\maW$, while the
singular values $\maSigma$ affect the effective subband channel
distortion $D_\calB$.

\section{Wideband Quantization} \mabel{sec:wq}
We consider a situation where multiple MIMO channels are operated in
parallel on subbands in the frequency domain, and the
frequency-selective channels are correlated between subbands as
discussed in Section~\ref{sec:sys}.

\subsection{Eigenvector Quantization}

In this section we consider wideband quantization pertaining to using
a budget of bits for feeding back information about the $K$
largest eigenvectors of a 
covariance matrix ${\maR}$ as precisely as possible. From Proposition~\ref{prop:overall}, the objective of wideband quantization
is to find 
\be
 \maW = \arg\min_{\bar\maW \in \calW} \, d_p(\bar\maW,\widetilde\maR)
\mabel{eq:wbcriterion}
\ee
where $d_p(\bar\maW,\widetilde\maR)$ is the projection distortion
(\ref{eq:projdist}), and $\calW$ is a codebook of $\Nt\times K$
matrices. As we are separately interested in the $K$ columns, Grassmannian codebooks~\cite{Ghosh2012,Schwarz19,Schwarz20,Schwarz22} of $K$-dimensional subspaces are not of interest. Also, due to the modular design in mind, the phases of the $K$ columns are irrelevant. Accordingly, codebooks of generic $N_t \times K$ unitary (Stiefel) matrices \cite{Schwarz2021} are not of interest either. Pertinent matrix codebooks would quantize the {\it flag manifold} ones~\cite{Pitaval2013}, i.e. Stiefel matrices modulo {\it scalar} column rotations. 

We stress that in contrast to the literature
\cite{Liu2017,Onggosanusi2018}, we use the sample covariance of the normalized subband channels $\tilh$.
According to Propostion~\ref{prop:overall}, this quantity
governs the contribution of wideband quantization to overall distortion. 

Motivated by Lemma~\ref{lem:isometry}, we want to feed back information about {\it orthonormal matrices}. Orthonormality can be guaranteed 
by directly
using a matrix codebook consisting of orthogonal
matrices~\cite{Ghosh2012,Schwarz19,Schwarz2021,Pitaval2013}.
%
To minimize the projection distortion, it is essential to have a high
granularity of $\calW$. If matrix codebooks of the kind of
\cite{Ghosh2012,Schwarz19,Schwarz2021,Pitaval2013} are used, high granularity leads to high complexity.
In~\cite{3GPP,Onggosanusi2018,Song2018}, complexity is kept at bay by using a
high-granularity Grassmannian vector codebook $\calV$ 
to quantize the eigenvectors one by one. The quantization process in~\cite{3GPP,Onggosanusi2018} 
proceeds as a series of parallel
exhaustive searches for the optimal codewords each of which quantizes
one column of $\maU_K$:
\begin{equation}
\vecv_k = \arg\underset{\vecv \in \calV}{\mathrm{min}} ~~  d(\vecu_k,\vecv) 
~~\mathrm{for}~k=1,2, \dots, K\,,
\mabel{eq:vecQuant}
\end{equation}
and the wideband feedback matrix becomes $\maV=[\vecv_1, \vecv_2,
  \dots, \vecv_K]$.
The codebook $\calV$ consists of $N_t$-dimensional unit norm vectors,
i.e. $\calV \triangleq \{\hat{\vecv}_1,\hat{\vecv}_2,
\dots,\hat{\vecv}_{2^B}\}$ and has cardinality $2^B$. The complexity
of finding a codeword using (\ref{eq:vecQuant}) is the $K$th root of 
the complexity of finding a codeword in matrix codebook $\calW$ with the same
granularity. 
Columnwise Grassmann line quantization has the added benefit that it
operates natively on the flag manifold, as opposed to a matrix Grassmann~\cite{Ghosh2012,Schwarz19,Schwarz20,Schwarz22} or Stiefel~\cite{Schwarz2021} quantization.  

However, if $\calV$ is overcomplete, 
there is a high
probability that $\maV \maV^\rmH \ne \maI_{N_t}$. According to
Proposition~\ref{lem:isometry}, this breaks the connection between
subband channel and effective channel quantization. Which is to say,
the basis spanned by a typical $\maV$ is probably independent but not
orthogonal,
which causes errors when quantizing the subband channel using $\maV$.
 
Motivated by this, we develop two wideband quantization
approaches, namely \textit{orthonormalized wideband precoding (OWP)}
and \textit{sequential wideband precoding (SWP)}, that improve not only
the quality of wideband feedback but also the accuracy of subband
channel projection---they enable high-granularity feedback of
orthogonal bases, with the same complexity as (\ref{eq:vecQuant}). 
   
\subsubsection{Orthonormalized Wideband Precoding (OWP)}\mabel{subs:owp}
We aim at $N_t \times K$ dimensional wideband feedback, using a vector
codebook $\calV$, and an independent quantization (\ref{eq:vecQuant})
of eigenvectors. The quantization data pertaining to the resulting
$\maV$  is fed back to the BS. 

It is important to distinguish between $\maV$ with columns in $\calV$, and the orthogonal basis matrix $\maW$ used to compute the subband feedback. We assume that the users and the BS
{\it have agreed on a method to orthonormalize} the vectors $\maV$.
There is  a deterministic function which produces an orthogonal
$\Nt\times K$ matrix from a generic $\Nt\times K$ matrix, e.g.
Gram-Schmidt orthogonalization;
\begin{equation} 
\maW=\mathrm{orth}(\maV)\,,
\end{equation}
which would recover precisely the result of the QR-decomposition in
(\ref{eq:wwo}). 
This method is
shared at the same time as the codebook $\calV$. Exactly the same
number of bits can thus be used to feed back the orthogonal
$\maW$ as feeding back the original $\maV$.

The user constructs the orthogonal $\maW$ and uses it to
compute the effective channels. The BS similarly constructs $\maW$ from $\maV$ and uses it as the basis in which subband feedback is interpreted. 

In \cite{Schwarz19}, scalar quantization is used to feed back information of a rectangular unitary matrix, and the recipient of the feedback extracts the unitary columns. The difference to OWP is that in OWP, both the user and the BS perform orthogonalization using {\it the same} orthogonalization algorithm, and the user uses the orthogonalized $\maW$ to determine and the BS to interpret subsequent feedback.   

\subsubsection{Sequential Wideband Precoding (SWP)}\mabel{sec:imp} 

In~\cite{Song2018}, a sequential method to find codewords $\vecv\in\calV$ was presented, with the objective to find the best basis $\maV$ of $K$ vectors for $N_t$-dimensional vectors $\vech$. \columnversion{We}{Here, we}
use a similar approach to find $\maV$ {\it and to orthogonalize it}. In \cite{Song2018}, the equivalent channel coordinates were computed by pseudoinversion, and the BS used $\maV$ when reconstructing the channel. In contrast, we compute the equivalent channel using the orthogonalized $\maW$, and the BS uses $\maW$; the users and the BS use the same orthogonalization principle.  

The intuition of SWP is
to provide  \columnversion{self-correction}{a self-correction capability}
to the quantization process. 
The quantization of eigenvector $k$ may
not only convey information about the eigenvector $\vecu_k$, it may also
convey information about the errors of quantizing the eigenvectors
$\vecu_j$ of higher eigenvalues $j<k$.

For this, define a projector $\maPi^\perp$ updated in each iteration $j$, and
initialized at the identity matrix. For convenience, we denote its value at
iteration $j$ as $\maPi^\perp,j$, and $\maPi^\perp_0={\bf I}_{N_t}$. 
This matrix characterizes the perpendicular space of all
previous feedback vectors.
In the $j$-th iteration we calculate the projection of the covariance
matrix $\widetilde\maR$ to the current perpendicular space $\maPi^\perp_{j-1}$:
\begin{equation}\mabel{seq-proj}
\widetilde\maR_j=\maPi^{\perp}_{j-1} ~\widetilde\maR~\maPi^\perp_{j-1},~j=1,\dots, K.
\end{equation}
Then the strongest eigenvalue eigenvector ${\bf e}_{j, 1}$ of $\widetilde\maR_j$
is found using, e.g., the power method.
Utilizing the vector quantization codebook we then quantize ${\bf
  e}_{j, 1}$ to $\vecv_{j}\in\calV$.

\begin{algorithm}[t]
\caption{Sequential Wideband Precoding (SWP)}
\mabel{alg:seq}
\begin{algorithmic}[1]
\STATE {\large I}nitialization  
\STATE ~~~~$\maPi^\perp_0={\bf I}_{N_t}$
\STATE  {\large W}ideband Quantization 
\STATE ~~~~\textbf{for} $j=1, 2, \dots, K$ \textbf{do} 
\STATE~~~~~~~Project $\widetilde\maR$ to $\maPi^\perp$ using \eqref{seq-proj}: $\widetilde\maR_j$ 
\STATE~~~~~~~Find strongest eigenvector ${\bf e}_{j, 1}$ of $\widetilde\maR_j$
\STATE~~~~~~~Quantize ${\bf e}_{j, 1}$ using wideband codebook
 to $\vecv_{j}$
\STATE~~~~~~~Project $\vecv_{j}$ to $\maPi^\perp$ with
\eqref{seq-orth}, normalize: $\vecw_{j}$
\STATE~~~~~~~Update $\maPi^\perp$ with $\vecw_{j}$ using  \eqref{seq-update}
\STATE~~~~\textbf{end for}
\STATE  {\large O}utput
\STATE ~~~~Obtain basis matrix $\maW =[\vecw_{1},\vecw_{2},\dots,\vecw_{K}]$
\STATE ~~~~Obtain feedback matrix $\maV=[\vecv_{1},\vecv_{2},\dots,\vecv_{K}]$
\end{algorithmic}
\end{algorithm}

Next, we project the quantized strongest eigenvector $\vecv_{j}$ to
the perpendicular space of the previous codewords:
\begin{equation}\mabel{seq-orth}
\tilde\vecv_{j}=\maPi^\perp_{j-1} \vecv_{j}, 
\end{equation}
and orthonormalize it to $\vecw_{j}=\tilde\vecv_{j,{\rm
    q}}/\| \tilde\vecv_{j} \|$.
Finally, 
 the perpendicular space is updated to 
\begin{equation}\mabel{seq-update}
\maPi^\perp_j=\maPi^\perp_{j-1}-\vecw_{j} \vecw_{j}^\rmH.
\end{equation}
Note that in each iteration, $\maPi^\perp_j$ is a projector;
$\left(\maPi^\perp_j\right)^2 = \maPi^\perp_j$ and
$\left(\maPi^\perp_j\right)^\rmH = \maPi^\perp_j$.

After $K$ iterations, the user feeds back \columnversion{}{the matrix}
$\maV=\left[\vecv_{1}, \vecv_{2},\dots,\vecv_{K}\right]$ consisting of
$K$ codewords in the
vector quantization codebook $\calV$ to the BS. The user applies the 
corresponding orthogonal matrix $\maW=\left[\vecw_{1},
  \vecw_{2},\dots, \vecw_{K}\right]$ when 
  deriving
subband feedback. The BS, knowing the SWP algorithm, can reproduce
$\maW$ from $\maV$, and uses it as the basis when interpreting
subband feedback.

SWP is summarized in Alg.\ref{alg:seq}. When considering this as an
algorithm for sequentially finding a $\maV$, there is a crucial
difference to \cite{Song2018}. In Alg.\ref{alg:seq}, projectors act on
the remainder, while in \cite{Song2018}, projectors act separately on
each codeword in $\calV$ at each stage of the algorithm. This results
in considerably higher computational complexity than
Alg.\ref{alg:seq}.

To formulate a performance measure, we need a technical characterization
of codebooks. We consider covariance matrices $\widetilde\maR$ with continuously
distributed eigenvectors, and a codebook $\calV$ quantizing vectors
$\vecu$ in this distribution to points $\vecv\in\calV$ that are at
most at chordal distance $r$ from $\vecu$. The quantization can be
characterized by the inner product $\vecu^\rmH\vecv$ of the normalized
vectors. If the probability distribution of inner products across all
$\vecu$ and their corresponding $\vecv$
does not depend on the phase of the inner product, and is a
non-increasing function of the chordal distance $d(\vecu,\vecv)$, we
call the quantization {\it radial}, and the chordal distance is the {\it
  quantization error}. For a more detailed discussion, see Appendix A.

Note that OWP of Section \ref{subs:owp} does not reduce the projection
distance as compared to using the possibly non-orthogonal $\maW$
produced by parallel quantization in (\ref{eq:vecQuant}). OWP only
produces effective channels which can be quantized based on isometry.
In contrast,
for SWP we have:

\begin{proposition}\mabel{theo:sequential}	 
Assume a radial codebook $\calV$ for quantizing the
distribution of eigenvectors of $\widetilde\maR$, with maximal quantization
error $r\leq1/\sqrt{2}$.
SWP
of Algorithm~\ref{alg:seq} provides a smaller projection distortion
than 
OWP of (\ref{eq:vecQuant}).
\end{proposition}

\begin{IEEEproof}
See Appendix A. 
\end{IEEEproof}
\smallskip

\subsection{Wideband Amplitude Quantization}

Quantizing the real-valued singular values $\maSigma$ is rather
straightforward. There is a budget of bits for feeding back
information about them, and scalar quantization can be directly used. 
With quantized  $\maV$, and especially with orthogonalized
$\maW$, the singular values computed from (\ref{eq:svd}) do not
characterize the relative weight of the columns in $\maW$ when
describing subbands. Instead we
calculate {\it wideband amplitudes} (\ref{eq:wbamp}) 
for each column $\mathbf{w}_j$ in $\maW$, and
feed back scalar quantized versions
$\{\hat{\sigma}_j\}$.

\section{Subband Effective Channel Quantization}\mabel{sec:sq}	

Now we concentrate on quantizing subband-specific effective channel coordinates. The BS and user share a $\Nt\times K$ wideband basis
matrix $\maW$ which reduces dimensionality of subband channels to $K$,
and the quantized wideband amplitude matrix $\hat{\maSigma}$.
%
The $K$-dimensional effective channel $\vecb_s = \maW^\rmH\vech_s$ becomes (\ref{eq:bss}) for perfect wideband feedback. It is not i.i.d., but according to
Lemma~\ref{lem:isometry} it is an isometry for orthogonal $\maW$.
The effective channel
$\vecc_s=\hat{\maSigma}^{-1} \vecb_s$
becomes the i.i.d. (\ref{eq:css}) for perfect wideband feedback. The squared chordal distance for the quantization is
\be
d^2(\vech_s,\hath) = 1 - \left|\vech_s^\rmH \hath \right|^2
= 1 - \left|\vecb_s^\rmH \hatb \right|^2
 = d^2(\hatSigma\,\vecc_s, \,\hatSigma\,\hatc)\,.
\mabel{eq:ChordalNt}  
\ee
Thus, if quantizing $\vecc_s$, we should use the quantization metric (\ref{eq:ChordalNt}) instead of chordal distance.

\subsection{Vector Quantization from Deformed i.i.d. Codebook}\label{sec:deformedIID}

Following~\cite{love2006limited,xia2006design}, an i.i.d. vector
quantization codebook $\calC$ may be
deformed with $\hat\maSigma$ to $\calB$ of (\ref{eq:calB}), which is
then used to quantize $\vecb_s$, minimizing (\ref{eq:ChordalNt}).
Even with a relatively small $K$, the precision required from the
i.i.d. codebook $\calC$ may be prohibitive for applying full-fledged
vector quantization. We shall use product codebooks
(PCB)~\cite{Yuan2012,au2011trellis,Choi2015} with a trellis structure
based on $N_\ell< K$ -dimensional component codes. With $N_b$ bits per
component, we use $N_t/N_\ell\times N_b$ bits in total. To deform the
quantization according to~\cite{love2006limited,xia2006design},
we incorporate
$\hat\maSigma$ to the trellis metric in a blockwise manner,
corresponding to the component code being processed.

\subsection{Scalar Quantization with Adaptive Bit Allocation}\label{sec:ScalarQ}
As an alternative, scalar quantization of the entries of $\vecc_s$ may
be applied, with adaptive bit-allocation derived from $\hat\maSigma$.
Further simplification can be achieved by separately quantizing phase
and amplitude~\cite{Dowhuszko2015}.
Scalar quantization has the benefit that large codebooks can be used
with limited quantization complexity, as quantization is coordinate by
coordinate.
The {\it normalized} subband
effective channel $\vecc$ is a $K \times 1$ i.i.d. vector $\vecc=[c_1,
  c_2, \dots, c_K]^\rmT$. As overall phase is irrelevant, one of the
coordinates acts as phase reference. The objective is to minimize
(\ref{eq:ChordalNt}).

\subsubsection{Bit Allocation in Extrinsic Order} 

In 5G NR~\cite{3GPP,Onggosanusi2018}, quantization bit allocation is
based on the order of
  singular values $\hat{\sigma_i}$. 
This ordering is {\it extrinsic} to the effective channel coordinates
$c_i$. The coordinate $c_{\max}$ with largest $\sigma_i$ is
used as phase and amplitude reference.
For $m$ the coordinates with the next largest $\sigma_j$, $c_j/c_{\max}$
are uniformly quantized in phase with $B_\ell$ bits and in amplitude
with one bit to $\{1,1/\sqrt{2}\}$. The power ratio in the codebook is
thus $\eta=2$. For the remaining $K-(m+1)$ coordinates,
amplitudes are quantized to $1/\sqrt{2}$ (no bits), while $B_s <
B_\ell$ bits are used for the phases. The number of bits used for
quantizing $\vecc$ is thus
$L=(B_\ell+1) m + B_s (K-m-1)$.

\subsubsection{Bit Allocation in Intrinsic Order}
Starting from first principles we develop an adaptive bit allocation
principle taking the sizes of $c_j$ into account. Accordingly, this
scheme is based on an \emph{intrinsic order}.
When using coordinate-specific codebooks $\calC_i$
divided to separate amplitude
and phase
parts, the crucial observations to make from
\columnversion{(\ref{eq:deconstruct},\ref{eq:ChordalNt})}{(\ref{eq:bsss},\ref{eq:ChordalNt})} are that
effective channel coordinates
  $c_i$ are {\it fading} $\mathcal{CN}(0,1)$ random variables; 
as they are i.i.d., their quantizations should have the same mean energy $E_{\hat c_i\in \calC_i} \left\{|\hat c_i|^2\right\} =
  a^2$; 
subject to this, the {\it weighted inner product}
  $
   \left|\vecc^\rmH\hatLambda\,\hatc\right| = \sum_{j=1}^K  \,
   \hat\sigma_j^2\,  |c_j^* \hat c_j|
  $
   should be maximized. The contribution of $\hat c_j$ to feedback
   error depends on the {\it weighted amplitude} $\hat\sigma_j |c_j|$.
Following these principles we first normalize $\vecc$, and then
quantize all amplitudes with one bit. Assuming perfect phase
quantization, for Rayleigh fading variables the optimum quantization
levels have a power ratio of $\eta\approx5$, see~\cite{Pearlman}.
Denoting the quantized amplitude of coordinate $c_i$ as $\hat a_i$, we
use $\alpha_i = \hat{\sigma}_i\, \hat a_i$ to order the coordinates
for phase bit allocation.
The coordinate with largest $\alpha_i$ is used as phase
reference. To the coordinates with the $m$ next largest $\alpha_i$, we
allocate $B_\ell$ phase bits, while to the $K-(m+1)$ coordinates with
smallest $\alpha_i$, we allocate $B_s$ phase bits. The total number of
subband quantization bits required becomes:
$
L=K+B_\ell\, m+B_s(K-m-1)
$.

	\begin{table*}
 	\begin{center}  
			\caption{Quantization methods simulated}  \mabel{QuantiOpts} 
		        \begin{tabular}{|c|l|l|c|c|c|}  \hline
Fig.         & Quant. problem &  distortion metric  & CB  & Benchmark \\
                          \hline
\ref{simW1}  & WB vector  & chordal & \red{TSODFT, DD} & TSODFT \\ \hline
\ref{simW2}  & WB matrix   & projection&  \red{TSODFT}, DD  & TSODFT  \\
             &    & & \red{OWP, SWP} & OWP \\
             & & & \red{${\bf  R}_{B00B}$}, ${\bf R}_{B+B-}, {\bf R}$ & $\maR$\\ \hline
\ref{simS1} & Subband & chordal &\textcolor{red}{PCB, INT5, EXT2} &  EXT2\\ \hline
\ref{simS2},\ref{fig:coro1} &  Overall&  chordal & IND, OWP, SWP & IND \\
            &  quantization &  & PCB, INT5, EXT2 & EXT2 \\ 
&         &  & TSODFT, ${\bf R}_{B00B}$ &  \\ \hline
\ref{simSE} & MU-MIMO ZF & Spec. Eff. & IND+EXT2, OWP+INT5, SWP+INT5 & IND+EXT2 \\
&         &  & TSODFT, ${\bf R}_{B00B}$ &  \\ \hline
			\end{tabular}  
 	\end{center} 
	\end{table*}

\section{Simulations}\mabel{sec:sim}

The simulation results are organized as follows:
Sec.\ref{secB}-Sec.\ref{secD} show the results of wideband, subband
and overall quantization in order, as summarized in
Table~\ref{QuantiOpts}. We compare the quantization performance of different methods and select superior methods (in red) for subsequent comparisons. 

\subsection{Settings and Schemes}

For simulating the developed modular CSI quantization scheme, we need
a channel model with a realistic combination of channel directivity
and frequency domain correlation, as well as a realistic model of an
mMIMO antenna array.

\subsubsection{Channel Model} \hfill

The BS has a Uniform
Planar Array (UPA) with $N_t=32$ antennas, divided into $8\times
2\times 2$
horizontal $\times$ vertical $\times$ polarization elements,  depicted in Fig.~\ref{UPA32}.
The center frequency is $1.84$ GHz and  the bandwidth is
$18$ MHz with  $1200$ subcarriers divided into
$S=30$ subbands.
We use QuaDRiGa V2.0.0 \cite{quadriga} to generate MIMO channels with 3GPP 38.901 UMa NLOS settings, assuming $8$ scattering clusters and a ratio of 80\% indoor users.

We model a 120-degree sector of a base station, and 1000
single-antenna users are dropped uniformly at random within 250 m from the BS in the sector. Users have omnidirectional antennas, and as we only take users in one sector, BS antennas are considered omnidirectional as
well. For each user, the covariance matrix $\widetilde\maR$ is found by
averaging over subbands.

\begin{figure}
\centering
\includegraphics[width=0.4\textwidth]{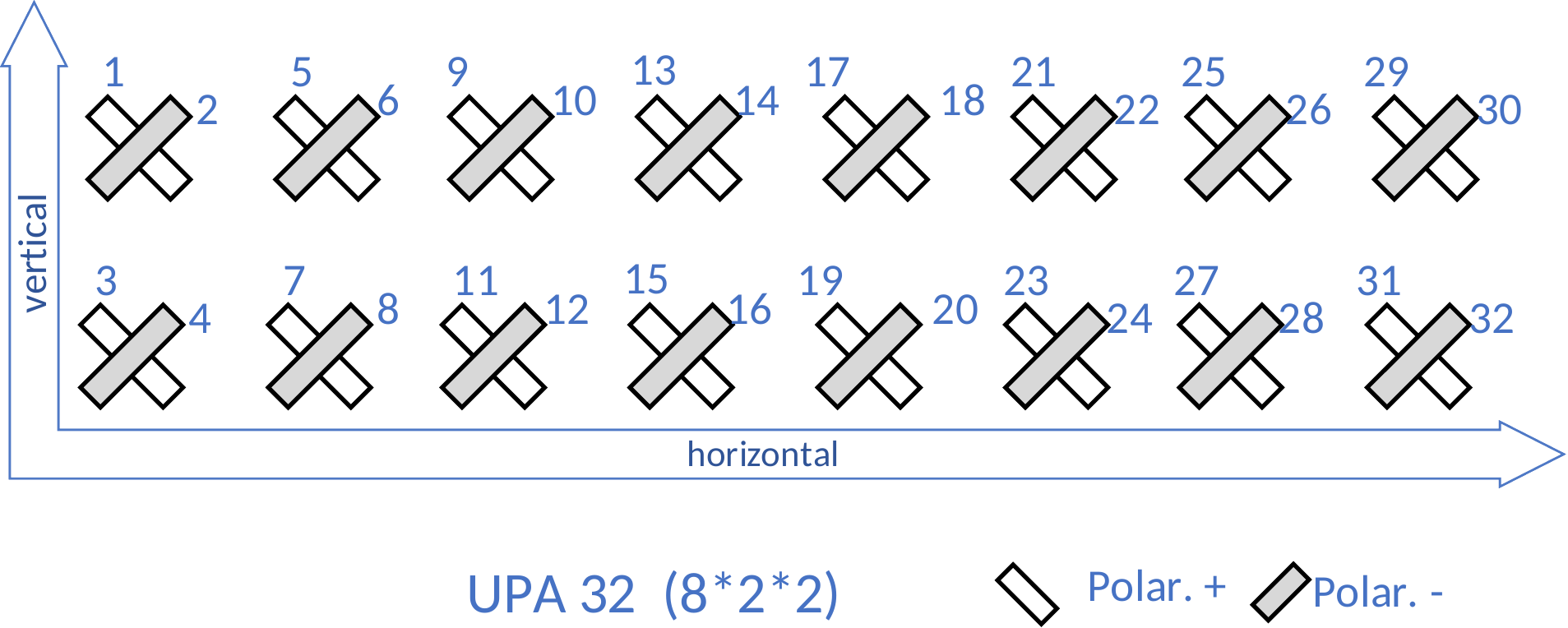}
\caption{($8\times 2\times 2$)-element UPA diagram illustration.}
\mabel{UPA32}
\end{figure}

\subsubsection{Polarization Decomposition of Channel} \hfill

With the UPA antenna arrangement of Fig.~\ref{UPA32}
the subband channel vector $\vech$
is a 3D Kronecker
product \cite{ying2014kronecker}: 
$$\vech=\vech_h \otimes \vech_v \otimes \vech_p\,.$$ 
Here $\vech_h,
\vech_v, \vech_p$ denote the horizontal, vertical and polarization
parts with $8, 2, 2$ elements, respectively.
The channel covariance matrix $\widetilde\maR$ and the wideband codebook for
$\maW$ are correspondingly decomposed.

Different polarizations of BS antennas are weakly correlated within a sector,
see~\cite{Hamalainen2003}. This can be used to reduce feedback
pertaining to $\widetilde\maR$.
We rearrange the channel vectors into $\vech^\rmT =
[\vech_+^\rmT,\, \vech_-^\rmT]$,  
according to the $+/-$ polarization. 
In addition to quantizing the \textit{fully-occupied} covariance
matrices, we consider two reduced alternatives assuming weak
cross-polarization correlation. The correlations of the
two polarizations are assumed either the same, or different:
\begin{equation}
   {\bf R}_{B+B-}=
                 \mat{ll}{\maB_+ &  \bf0\\ \bf0 & \maB_-}
   ~{\rm or}~~
   {\bf R}_{B00B} = 
     \mat{cc}{\maB &  \bf0\\ \bf0 & \maB}\,.
\label{eq:BOOB}
\end{equation}
Here $\maB$, $\maB_+$ and $\maB_-$ are $N_t/2 \times K/2$ matrices,
and $\maB = (\maB_+ + \maB_-)/2$. In simulations,
the $K=8$ strongest eigenvectors of $\widetilde\maR$ are used, while for $\maB$,
$\maB_+$, or $\maB_-$, the $K/2 = 4$ strongest are quantized.

\subsubsection{Wideband Quantization}\mabel{subs:schemes}
For quantizing the covariances arising in the three polarization
partition methods above, we apply
\begin{itemize}
\setlength\itemsep{0em}
\item IND: the standardized method used in \cite{3GPP} where {\it wideband vectors are independently quantized};
$\maW$ and $\hatSigma$ are direct quantizations of $\maU$ and $\maSigma$ from \eqref{eq:svd}. 
\item OWP: $\maW$ is orthogonalized, wideband amplitudes are
    \eqref{eq:wbamp} and effective channels (\ref{eq:deconstruct}).

  \item SWP: $\maW$ is orthogonalized with  Alg.\ref{alg:seq}.
\end{itemize}
 
The singular values or wideband amplitudes $\{\sigma_j\}$ are scalar
quantized. The strongest beam is indicated and the amplitudes of the
remaining $K-1$ beams are quantized by $3$ bits each using the
codebook $\{1/\sqrt{2^{m}} \}_{m=0}^6 \cup \{0\}$ from \cite{3GPP}. 

\subsubsection{Wideband Vector Quantization Codebooks}

Following~\cite{Onggosanusi2018,Song2018} we apply
array architecture aware precoding for UPA.  We decompose wideband 
vector codebooks as $\calV_h \otimes \calV_v \otimes \calV_p$, where
$\calV_h, \calV_v$ and $\calV_p$ quantize the horizontal, vertical and
polarization dimensions, respectively. We use oversampled DFT
codebooks for $\calV_h$ and $\calV_v$ in respective dimensions,
resulting in a Tensored Oversampled DFT (TSODFT) codebook. In the
fully-occupied case, for $\calV_p$ we use a $2$-bit binary chirp (BC)
codebook~\cite{Howard2008}, which is an ideal i.i.d. codebook of this
size.

We also use data-driven (DD) Grassmann quantization codebooks, found
by applying the Lloyd algorithm to a training set of eigenvectors of
wideband covariance matrices. Compared with random vector codebooks,
data-driven Grassmann quantization codebooks take the distribution of
data samples to be quantized into account, reducing distortion.

\subsubsection{Subband Quantization Codebooks}
For subband feedback, we use the options 
\begin{itemize}
\setlength\itemsep{0em}
\item Correlated vector quantization based on a product codebook,
  denoted as PCB
\item Adaptive  scalar quantization with extrinsic order
  and power ratio $\eta=2$, denoted as EXT2
 \item Adaptive scalar quantization with intrinsic order and power
   ratio $\eta=5$, denoted as INT5
\end{itemize}
PCB is discussed in Section \ref{sec:deformedIID}, and the scalar
quantizations in Section~\ref{sec:ScalarQ}.

\begin{figure*}[t]
  \centering
  \begin{minipage}[b]{0.45\textwidth}
    \includegraphics[width=\textwidth]{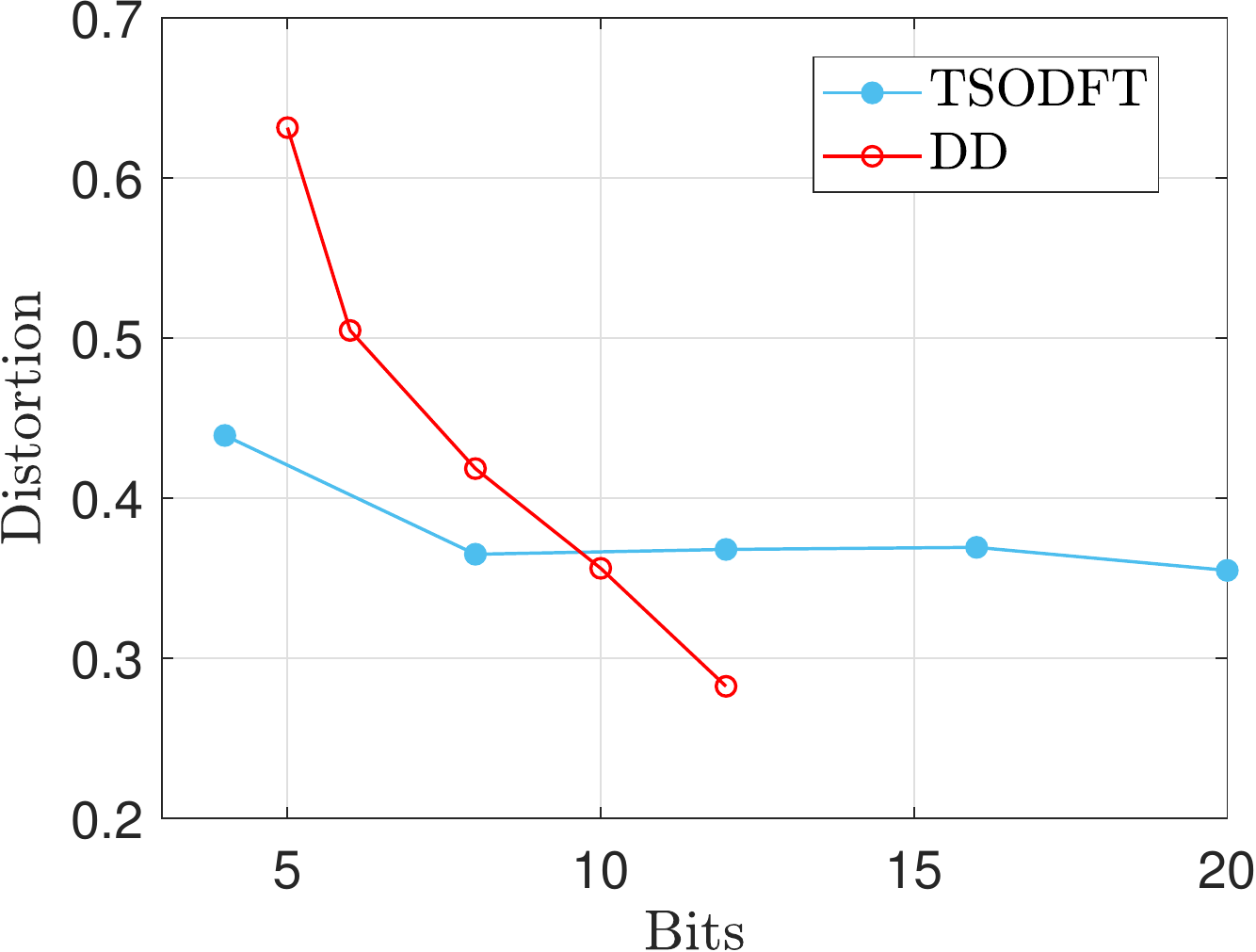}
    \caption{Wideband vector codebook distortion. Tensored Oversampled DFT (TSODFT) vs. Data Driven (DD) codebooks.}
    \mabel{simW1}
  \end{minipage}
  \begin{minipage}[b]{0.08\textwidth}
\phantom{5}
\end{minipage}
  \begin{minipage}[b]{0.45\textwidth}
    \includegraphics[width=\textwidth]{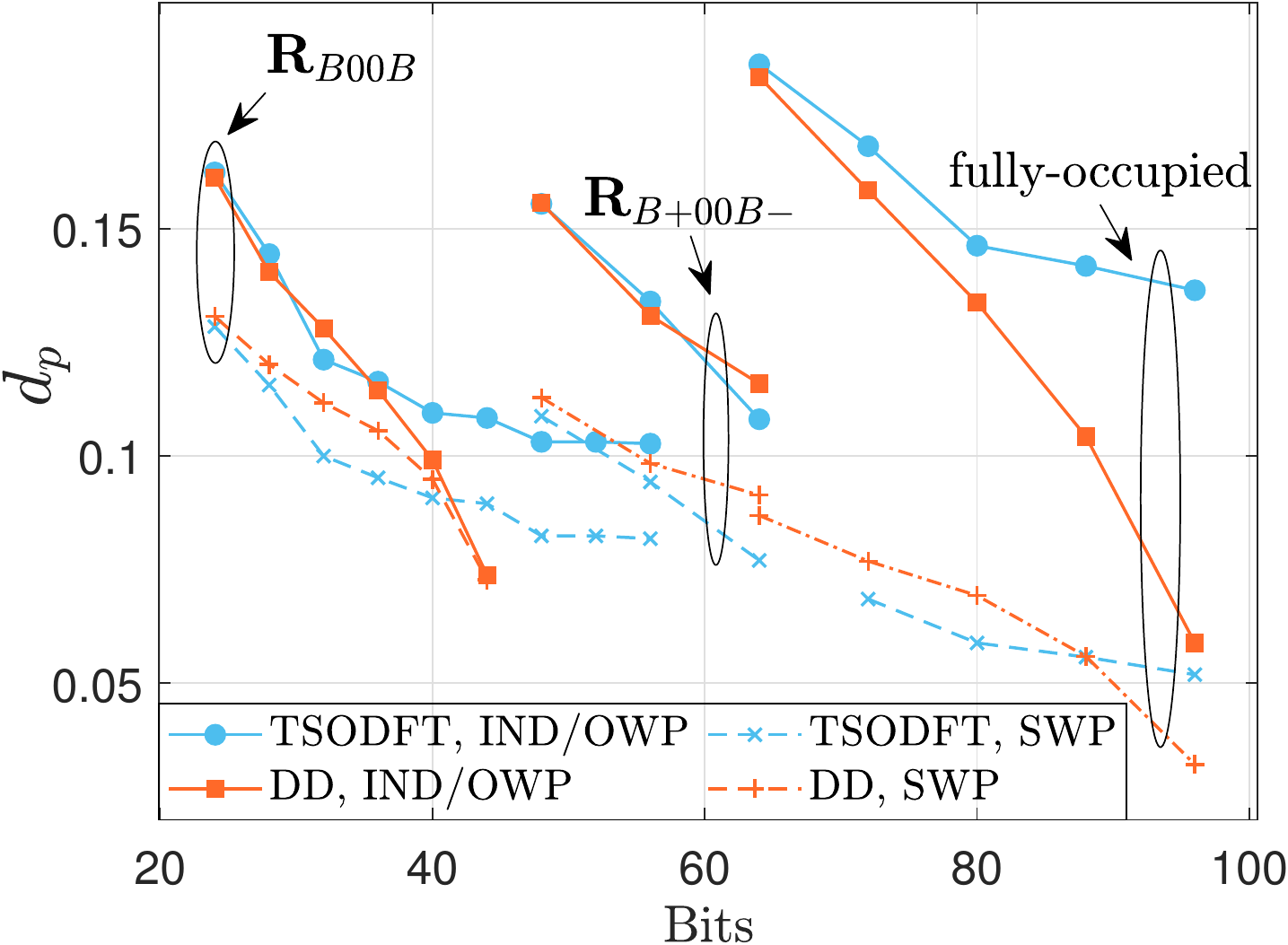} 
    \caption{Wideband \columnversion{}{matrix} projection distortion. Four combinations of codebooks (TSODFT or DD) and matrix quantization principles (OWP or SWP) for different dolarization decompositions.}
    \mabel{simW2}
  \end{minipage}
\end{figure*}
 
\subsection{Wideband Quantization Performance Comparison}\label{secB}

\subsubsection{Vector Quantization Distortion w.r.t. $\vecu_k$}
We first compare the wideband 
vector quantization codebooks used for quantizing the $K=8$ strongest eigenvectors $\vecu_k$ of sample covariance matrices. In Fig. \ref{simW1} we see that with smaller numbers of bits, the array architecture aware TSODFT codebook performs better than the data driven one, while with increasing numbers of bits, the DD codebook is able to learn the distribution of the eigenvectors, resulting in smaller distortion. 

\subsubsection{Projection Distortion w.r.t. $\widetilde\maR$}
We compare OWP and SWP w.r.t. the projection distortion  \eqref{eq:wbcriterion} under three polarization structures, ${\bf R}_{B00B}$, ${\bf R}_{B+B-}$ od \eqref{eq:BOOB}, and the fully-occupied $\bf R$, for TSODFT and DD codebooks.

In Fig.~\ref{simW2}, each of the twelve combinations is simulated for several codebook sizes, and projection distortion is reported against the total number of 
bits.\footnote{For clarity, the overlapping points from ${\bf R}_{B+B-}$ are removed. The product of the TSODFT
codebook over-sampling ratios in horizontal and vertical dimensions is $\omega=2^n, n \in [2,9]$. The best division of oversampling between horizontal and vertical is applied. The total number of bits
used for feeding back $\maW$ is $\log_2(\omega\, N_t/2)\,
K/2$ for $\maR_{B00B}$, $\log_2(\omega\, N_t/2)\, K$ for $\maR_{B+B-}$
and $\log_2(\omega\, N_t)\, K$ for the fully-occupied case.
}
Recalling that the projection distortion $d_p$ does not depend on orthogonalization, the same distortion arises for OWP and independent vector quantization IND, while SWP may have different $d_p$. In Fig.~\ref{simW2}, 
SWP outperforms IND/OWP for all polarization
structures, which agrees with Proposition~\ref{theo:sequential}. In most cases the gap between SWP and IND/OWP is large. When the number of bits increases, saturation can be observed for TSODFT, in accordance with Fig. \ref{simW1}. Interestingly, when the number of bits increases, OWP and SWP converge for DD. The reason is that when the data driven approach starts to learn an appropriate quantization for the eigenvectors of the channels, the gain provided by SWP in correcting quantization errors for the largest eigenvectors reduces. 

Polarization structure ${\bf R}_{B00B}$ provides the
best trade-off between performance and codebook cardinality.
Fully-occupied SWP yields the lowest 
$d_p$, at the cost of twice the number of bits.
We focus on ${\bf R}_{B00B}$ and TSODFT below due to easy implementation and good performance.

\subsection{Subband Quantization Performance Comparison}\label{secC}

We compare subband quantization schemes in terms of the distortion $D_{\calB}$ of subband effective channel quantization. 
Here, a subtle effect comes into play. A given $\widetilde\maR$ has singular values $\maSigma$, with a spread $\sigma_{\max}-\sigma_{\min}$. The correlation structure $\hat\maSigma$ of subband quantization, however, depends on the wideband quantization scheme. With wideband quantization precision decreasing, the spread of $\hat\maSigma$ decreases and the volume of the subband effective channel space $\maW^\rmH\vech_s$ and the corresponding distortion $D_{\calB}$ increases. 
In Fig. \ref{simS1} we consider $D_\calB$ with {\it ideal} wideband quantization and with OWP. 
EXT2 and INT5 based on scalar quantization with bit allocation, as well as correlated PCB are 
considered. \footnote{For PCB, we use between $10$ and $50$ bits per subband, adjusting the values of $N_b$ and $N_l$. For EXT2 and INT5, the number of strong beams $m$ varies from $0$ to $7$ with phase bit allocations $(B_l,B_s) \in \{(3,2),(4,3),(5,4),(6,5)\}$.} PCB performs the best followed by INT5 while EXT2 is the worst. Hence, we focus more on comparing INT5 and PCB below. In OWP with smaller singular value spread than with ideal wideband quantization, INT5 almost catches up with PCB. For SWP, the $D_\calB$-values would be closer to the ideal case, while for IND, considering $D_\calB$ does not make sense as the independently quantized $\maV$ is not an isometry of chordal distance. 

\begin{figure*}[t]
  \centering
  \begin{minipage}[b]{0.45\textwidth}
  \includegraphics[width=\textwidth]{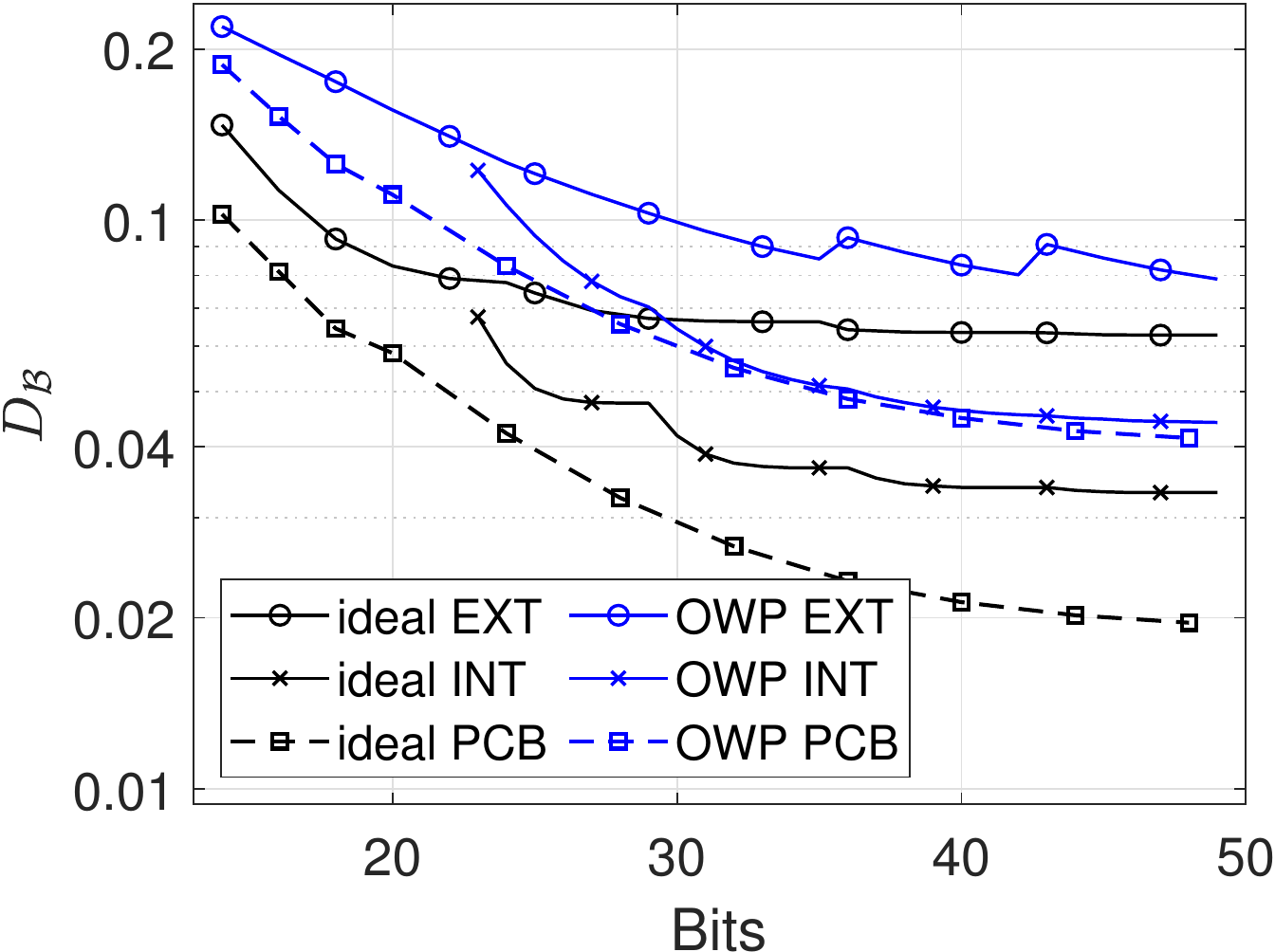}
    \caption{Subband codebook distortion with ideal wideband quantization, and with OWP, for subband codebooks EXT, INT and PCB.}
    \mabel{simS1}
  \end{minipage}
  \begin{minipage}[b]{0.08\textwidth}
\phantom{5}
\end{minipage}
   \begin{minipage}[b]{0.45\textwidth}
    \includegraphics[width=\textwidth]{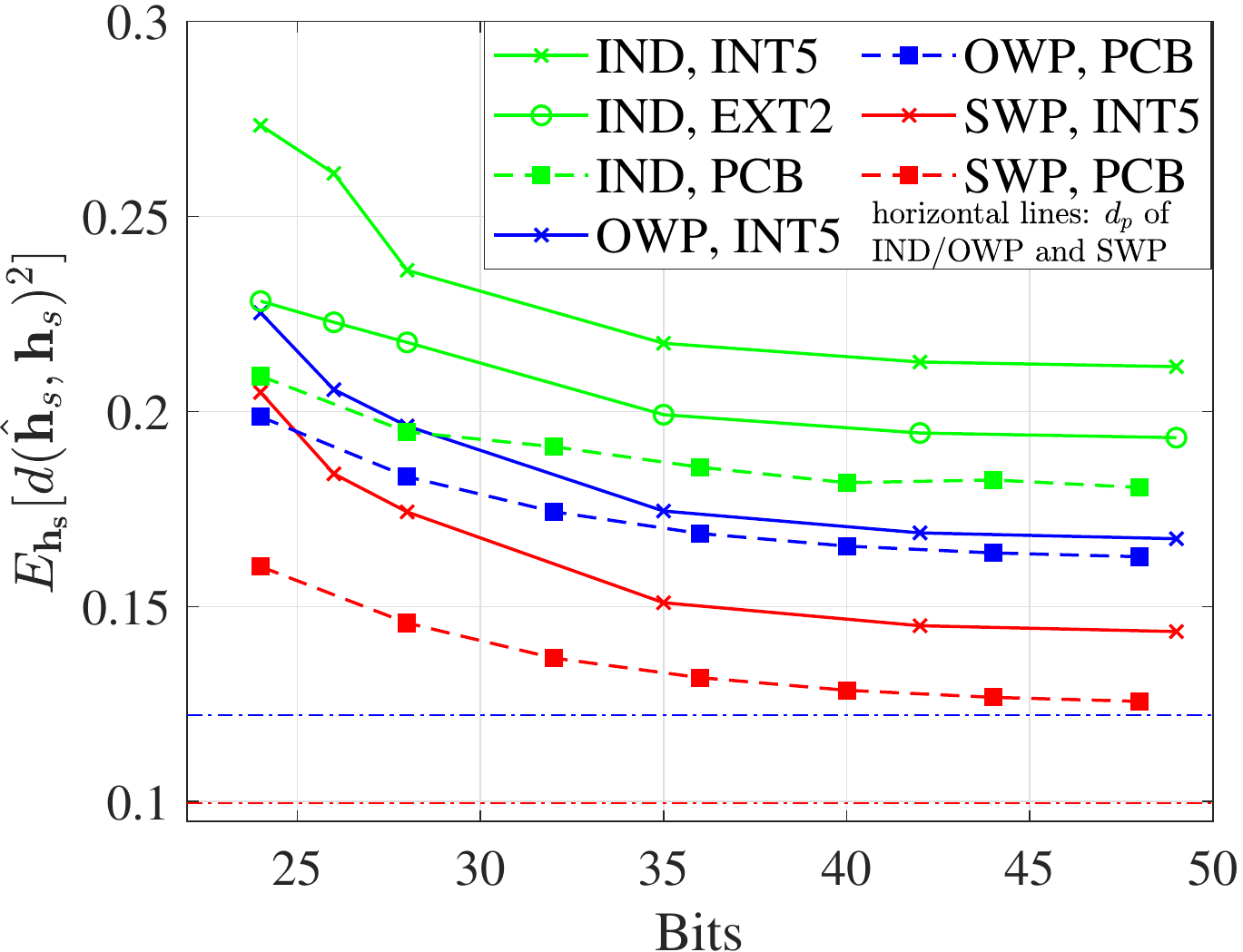}
    \caption{Overall distortion comparison. Combinations of wideband matrix quantization principles (IND,OWP,SWP) and subband codebooks (INT5,EXT2,PCB) for WB vector codebook TSODFT, and  polarization structure ${\bf R}_{B00B}$.}
    \mabel{simS2}
  \end{minipage}
\end{figure*}

\begin{figure*}[t]
  \begin{minipage}[b]{0.45\textwidth}
  \includegraphics[width=\textwidth]{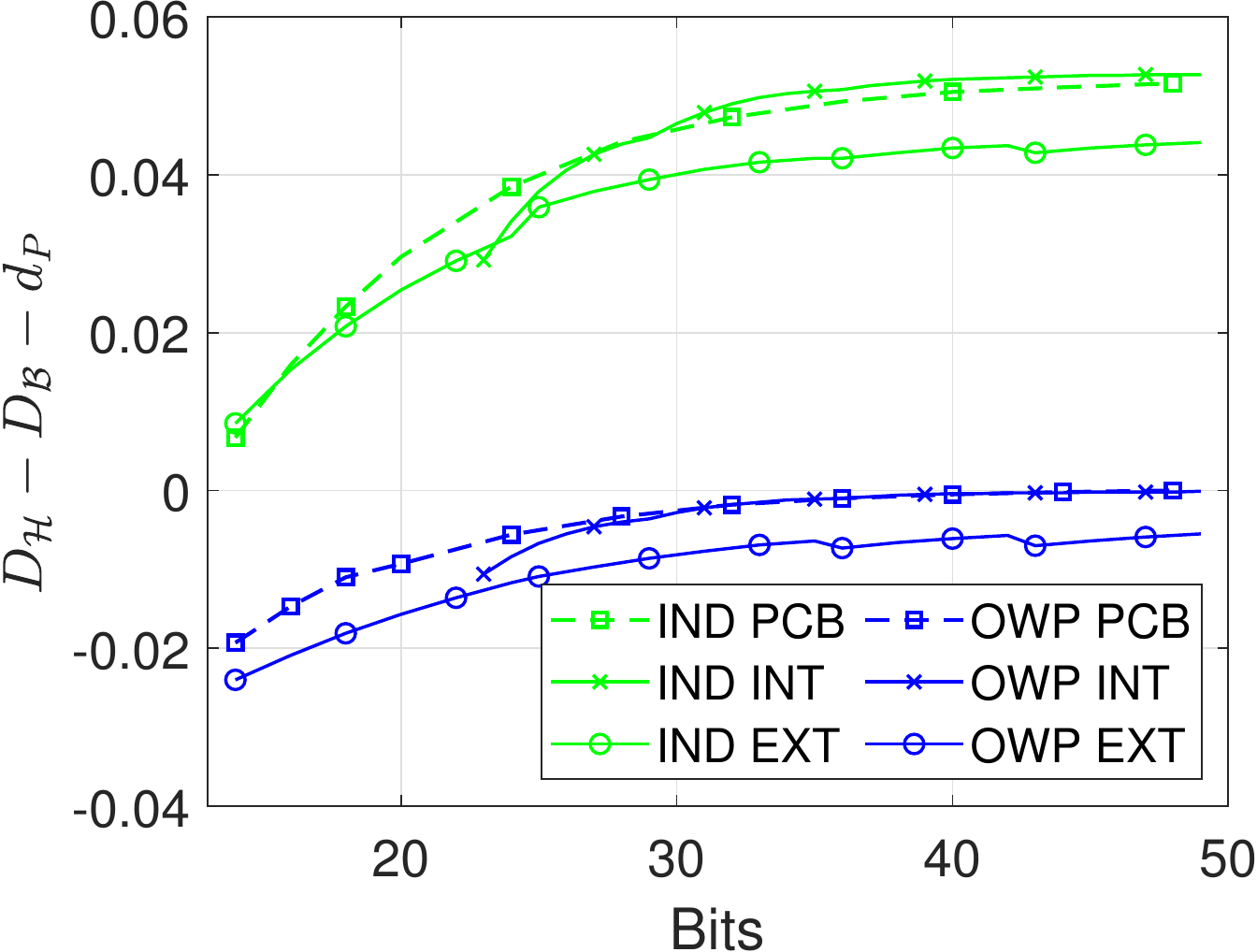}
  \caption{Difference between total distortion $D_\calH$ and upper bound $D_\calB + d_p$ of Corollary~\ref{cor:bounds}, valid for orthogonal \columnversion{WB}{wideband} feedback (OWP).}
  \mabel{fig:coro1}
\end{minipage}
\begin{minipage}[b]{0.1\textwidth}
\phantom{5}
\end{minipage}
\begin{minipage}[b]{0.45\textwidth}
\includegraphics[width=0.95\textwidth]{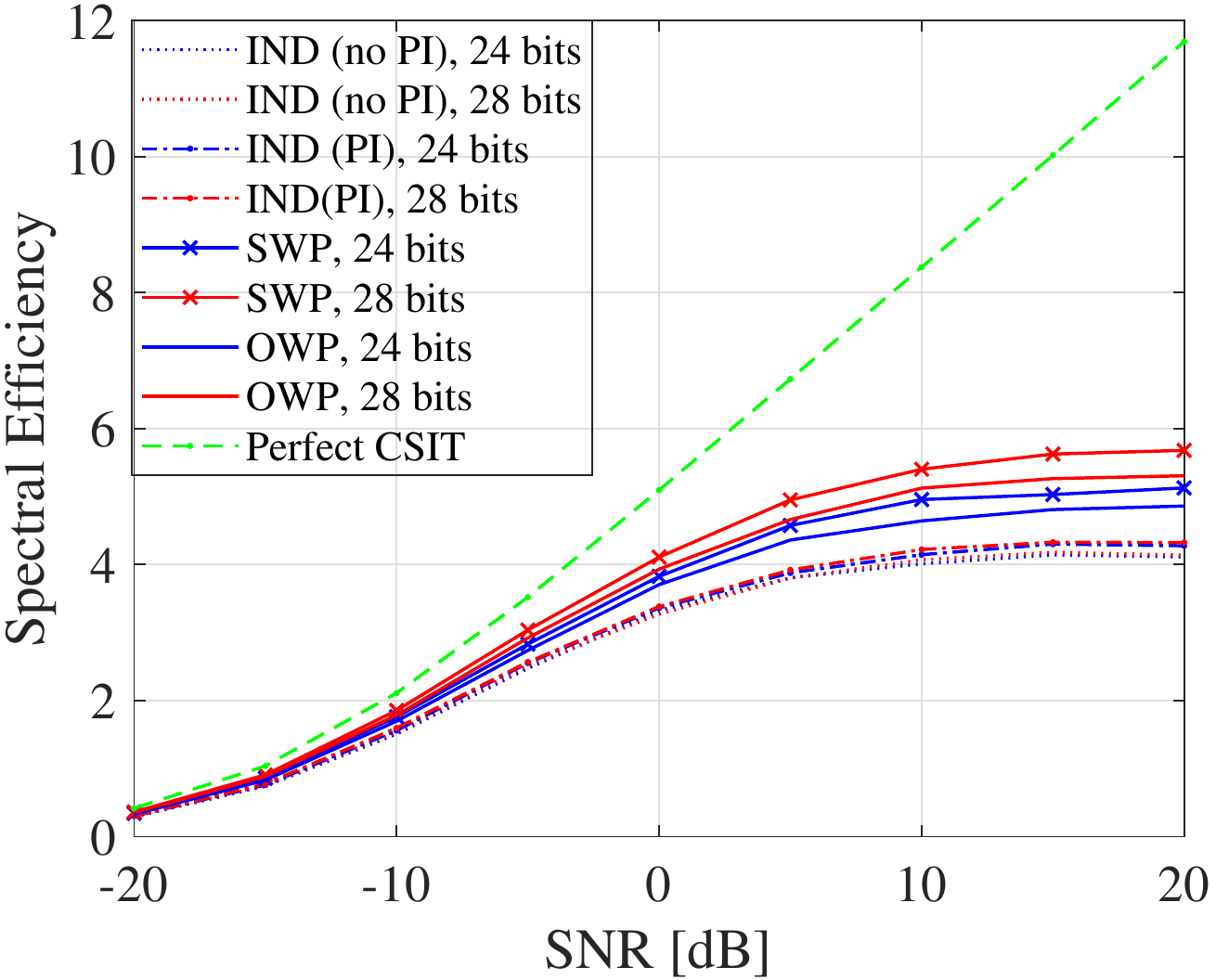}
\caption{Multiuser scenario: Spectral efficiency with 4 users: SWP, OWP and IND with/without pseudoinversion (PI).}
\mabel{simSE}
\end{minipage}		
\end{figure*}

\subsection{Overall Quantization Distortion}\label{secD}
Next we consider the overall quantization distortion
$D_\calH  = E_{\vech_s}[d(\hath_s,\vech_s)^2]$. 
For OWP and SWP wideband feedback we consider INT5 
and PCB, according to 
Fig.~\ref{simS1}. For independent wideband quantization we 
consider all three  subband quantization schemes, EXT2,  
INT5 and  PCB.
TSODFT with over-sampling ratio $\omega=16$ is used for
polarization structure ${\bf R}_{B00B}$. 

The resulting overall distortion vs. the number of bits for subband quantization is reported in Fig.~\ref{simS2}.\footnote{Here independent wideband quantization uses pseudoinversion for deriving effective channels which gives higher accuracy quantization.
Feedback sizes $L \in\{ 24,26,28,35,42,49\}$ bits per subband coordinate are
considered. For these, we have $(B_l,B_s)=(3,2)$ and $m=5,6,7$ for EXT2
and $m=2,4,6$ for INT5. In addition we have $m=7$ for EXT2 and $m=6$
for INT5, and $(B_l, B_s) \in \{(4,3),(5,4),(6,5)\}$, c.f. Section~\ref{sec:ScalarQ}.
PCB is generated with $N_l=2$ and $N_b=6,\ldots, 12$. }
For OWP, the performance difference between PCB and INT5 is fully captured by the subband performance in Fig.~\ref{simS1},  
in concordance with Corollary~\ref{cor:bounds}. The gain of PCB is enhanced when combined with SWP, and for both OWP and SWP, the total distortion is lower bounded by the corresponding projection distortion, in line with Proposition~\ref{prop:overall} and Corollary~\ref{cor:bounds}. 
For OWP and SWP are $d_p=0.122$ and $0.01$, respectively, plotted with dashed lines. 
For IND, EXT2 outperforms IND5, as expected from \cite{3GPP}. PCB, however, outperforms EXT2 also for IND.

To verify the validity of the upper bound in Corollary~\ref{cor:bounds}, we have plotted the difference between the overall subband channel distortion $D_\calH$ and the upper bound $D_\calB + d_p$ in Fig.~\ref{fig:coro1}, for OWP and IND. Subband effective channel distortion $D_\calB$ is calculated for the actual distribution of subband channels arising for a given wideband channel quantization, e.g., for OWP with PCB and INT, the $D_\calB$ from Fig.~\ref{simS1} are used. As OWP is based on orthogonal wideband feedback,  Corollary~\ref{cor:bounds} holds, and accordingly we see that the difference is $\leq 0$. For IND, the corollary does not hold---we see that {\it for non-orthogonal wideband feedback, the channel distortion $D_\calH$  may be much larger than what would be expected based on the part of the channel projected out, $d_P$, and the subband effective channel distortion, $D_\calB$.   
}

\subsection{Multiuser Spectral Efficiency}

To understand the effects of finite granularity modular CSI feedback on communication rate, 
we evaluate performance in a multiuser scenario
with $M=4$ single antenna users. The subcarriers are divided into $S=25$
subbands each with $48$ subcarriers. All users are simultaneously
served in the subbands.
Based on CSI fed back from the individual users, the BS performs zero
forcing (ZF) precoding on each subband. The three wideband
quantization schemes from Section \ref{subs:schemes} are compared,
with their respective best subband quantization with 24 and 28
bits.\footnote{To quantize $\vecc$, we use  parameters  $(B_l,
  B_s)=(3,2)$ and $m=5/7$ (IND) or $m=2/6$ (OWP and SWP) to get
  $24/28$ bits.} 
All users are assumed to have the same SNR. As a benchmark, ZF spectral efficiency with perfect channel state information at Tx is considered. 

Average single user spectral efficiency 
$R=\log_2(1+\gamma)$ with  $\gamma$ the ZF-SINR of a user is plotted vs. SNR in Fig.~\ref{simSE}.
Simulations corroborate the theoretical principles
discussed of sections \ref{ss:distortion} and \ref{sec:wq}. OWP
outperforms independent quantization by more than 25\%, while SWP
outperforms OWP by more than 8\%. Only nominal gain is achieved if users apply pseudoinversion to compute the coordinates with independent quantization, while the BS uses $\maV$. Increasing the effective
channel quantization granularity improves OWP and SWP considerably,
while providing little gain for independent quantization, confirming the distortion result in Fig.~\ref{simS2}.

\section{Conclusions}\mabel{sec:con} 
We have addressed modular CSI quantization for massive MIMO
systems. Analyzing the separation of feedback to a dimensionality reducing wideband part  and a lower-dimensional subband
part, we found quantization objectives for optimal modular
quantization. We show considerable performance improvement in a mMIMO
scenario when applying these principles. An orthonormalized wideband
precoding (OWP) scheme and a sequential wideband precoding (SWP)
scheme are developed.
Sharing a wideband feedback orthogonalization principle between the transmitter and receiver makes it possible to accurately quantize the lower-dimensional subbands. Without shared orthogonalization, improved subband quantization is not guaranteed to improve overall quantization accuracy.

It is worth noting that our central result, Proposition 1, holds for any situation where processing of a  vector is divided to two stages, where in the first stage the vector is dimensionality reduced to a subspace, and where the quality of the processing is measured by the chordal distance. Accordingly, this result readily generalizes to CSI feedback in a situation with multiple Rx antennas. This also extends the applicability of our results to mainstream Time Division Duplexing mMIMO, e.g. to front-hauled distributed processing for mMIMO, especially in the cell-free mMIMO case. 

\section*{Appendix}
\subsection{Proof of Proposition~\ref{theo:sequential}}

We first prove a needed lemma, and its corollary. We investigate
quantization codebooks $\calH$ in the space of unit norm $\Nt$-dimensional
complex vectors $\vecu$ modulo phase, i.e., the Grassmann manifold
$\calG_{\CC}(\Nt,1)$.
The chordal distance (\ref{eq:chordal}) is a metric on the
Grassmannian. For properties of metric balls, as well as geometric and
coding properties of the Grassmannian, see~\cite{Dai2008,Pitaval2018}.

The vectors $\vecu$ are quantized to a $\vecv\in\calV$ belonging to
the Voronoi cell $\Voronoi(\vecv)$. A metric ball $B(\vecv,r)$ with
radius $r$ centered at $\vecv$
consists of all $\vecu$ with $d(\vecu,\vecv)\leq r$.
To simplify the analysis we assume {\it radially distributed
  quantization errors}; the probability density of 
quantization error vectors
of an arbitrary $\vecu$ is rotationally invariant
and non-increasing in error magnitude. This means that 
the direction and magnitude of the quantization error are
independent random variables, and the distribution may be, e.g., uniform.
There is a maximum quantization error given by $\calH$,
s.t. the quantization errors are distributed in a ball $B(\vecu,r)$.
We have:

\smallskip
\begin{lemma}\mabel{lem:errorDistr}
Consider quantization of a source vector $\vecu\in
\calG_{\CC}(N,1)$ to a $\vecv$ radially
distributed in the ball $B(\vecu,r)$. Projecting and normalizing
$\vecu$ to an $m$-dimensional subspace yields $\vecu_{\rm s}$,
with
$\|\vecu^\rmH\vecu_{\rm s}\|^2 = {1-\eps^2}$ for an
$\eps \leq \sqrt{1-r^2}$,  while projecting and normalizing
$\vecv$
results in $\vecv_{\rm s}$.
The normalized projection $\vecv_{\rm s}$ is radially distributed in a
ball
with radius $r/\sqrt{1-\eps^2}$, centered at $\vecu_{\rm s}$, and with
the mean $\vecu_{\rm s}$. Denoting the quantization error after
projection as $d = d(\vecu_{\rm s},\vecv_{\rm
  s})$,  with the conditional distribution $f_{D|E}(d|\epsilon)$,
for any  $\eps>0$ there is a crossover
quantization error $d_c$ 
such that $f_{D|E}(d|0) > f_{D|E}(d|\epsilon)$ for $d<d_c$, and
$f_{D|E}(d|0) < f_{D|E}(d|\epsilon)$ for $d>d_c$.
\end{lemma}

\begin{IEEEproof}
Vector $\vecu$ is fixed. Take a point $\vecv \in B(\vecu,r)$. As we
are interested in chordal distance, all vectors are Grassmannian, i.e.,
they are equivalent under overall phase rotations. We decompose
$\vecu$ and $\vecv$ to components in the projection subspace and its
complement as
\bea
\columnversion{
  \vecu = \eps\, \vecu_{\bf o} + \sqrt{1-\eps^2}\, \vecu_{\rm s} \,,~~~
 \vecv = \alpha\, \vecv_{\bf o} + \sqrt{1-\alpha^2}\, \vecv_{\rm s} \mabel{eq:vdecomp}
}{
 \vecu &=& \eps\, \vecu_{\bf o} + \sqrt{1-\eps^2}\, \vecu_{\rm s} \mabel{eq:udecomp}\\
 \vecv &=& \alpha\, \vecv_{\bf o} + \sqrt{1-\alpha^2}\, \vecv_{\rm s} \mabel{eq:vdecomp}
}
\eea
Here, $\vecu_{\rm s}$, $\vecu_{\bf o}$, $\vecv_{\rm s}$ and $\vecv_{\bf o}$
are all unit norm vectors\columnversion{.}{, and $\epsilon$ and
  $\alpha$ represent the fraction of the vectors projected out.}

Now fix $\vecv_{\rm s}$ with $\left|\vecu_{\rm s}^\rmH\vecv_{\rm
  s}\right| = \cos\theta$ for some $\theta\in[0,\pi/2]$. For each
$\vecv_{\rm s}$ there is a preimage of vectors $\vecv\in B(\vecu,r)$
that project to $\vecv_{\rm s}$. It follows from the Cauchy-Schwarz
inequality that 
\be
\vecb = \frac1{\nu}\left(\eps\,\vecu_{\bf o} + \sqrt{1-\eps^2}\,\cos\theta\, \vecv_{\rm s}\right)
\mabel{eq:bN}
\ee
is the closest vector to $\vecu$ in this preimage. Here the
normalization is
\columnversion{
  $\nu = \sqrt{\eps^2 +  \cos^2\!\theta(1-\eps^2)}$.
}{
  \be
 \nu = \sqrt{\eps^2 +  \cos^2\!\theta(1-\eps^2)}\,.
 \ee
}
The preimage isomorphic to a set of unit vectors up to overall phase
in $N-m+1$ dimensions, i.e. $\calG_\CC(N-m+1,1)$. For concreteness,
the basis vectors can be taken as $\vecu_{\rm o},\vecv_{\rm s}$
together with $N-m-2$ null-vectors of the projection orthogonal to
$\vecu_{\rm o}$. The projection maps all of these vectors to
$\vecv_{\rm s}$. All vectors $\veca$ in the preimage should have
$|\vecu^\rmH\veca|^2 \geq 1-r^2$. It follows again from Cauchy-Schwarz
that the vectors in the preimage with largest chordal distance to
$\vecb$ are $\veca_{\max} = \beta \vecb + \vecb_{\rm o}$, where $\beta
= \sqrt{(1-r^2)/\nu^2}$, and $\vecb_{\rm_o}$ is orthogonal to
$\vecu_{\rm o}$ and $\vecv_{\rm s}$. These vectors are at chordal
distance
\columnversion{ $r_\nu = \sqrt{1-(1-r^2)/\nu^2}$ }{
\be
r_\nu = \sqrt{1-(1-r^2)/\nu^2}
\mabel{eq:rnu}
\ee}
from $\vecb$. The preimage is thus an $N-m+1$-dimensional ball
$B(\vecb,r_\nu)$ centered at $\vecb$. 
It is easy to convince oneself that the preimages of all $\vecv_{\rm s}$
are distinct, recalling that the overall phases of both $\vecv$ and
$\vecv_{\rm s}$
are irrelevant.
 
For a given $\epsilon$ there exists a smallest $\theta$ for which the
ball has non-negative radius, $\theta_{\min} = \arccos \sqrt{
  (1-\eps^2-r^2)/(1-\eps^2)}$, leading to a maximal chordal
distance between  projections: 
\be
 d^2(\vecu_{\rm s},\vecv_{\rm s}) \leq r_{{\rm s}} = \frac{r}{\sqrt{1-\eps^2}}~.
\mabel{eq:rperp}
\ee

\columnversion{}{\par}
The probability density at $\vecv_{\rm s}$ is an integral over the
preimage $B(\vecb,r_\nu)$. 
For any $\vecv_{\rm s}$ at fixed distance from $\vecu_{\rm s}$,
i.e. fixed $\theta$, the geometry and the probability density in the preimage are the same.
The probability of $\vecv_{\rm s}$ thus only depends on
$d(\vecu_{\rm s},\vecv_{\rm s})$; we have $f(\vecv_{\rm s}) \sim
f_{\Theta|E}\left(\arccos \left(|\vecv_ {\rm s}^\rmH\vecu_{\rm s}|\right)|\eps\right)$
for some $f_{\Theta|E}(\theta|\eps)$.

Now, from (\ref{eq:bN}) and \columnversion{the expression of
  $r_\nu$}{(\ref{eq:rnu}) we see that}, for fixed $\eps$ the radius of
$B(\vecb,r_\nu)$ decreases with $\theta$, while the distance of $\vecb$ from
$\vecu_{\rm o}$ grows. 
The ball always has a point with maximum chordal distance $r$ from
$\vecu$, which has the smallest probability density, while the
probability density at the center, $\vecb$, decreases with $\theta$.
Furthermore, the integration measure in
the preimage decreases with increasing $\alpha$ as
$\sqrt{1-\alpha^2}$.
The attenuation due to the integration measure thus increases with distance from
$\vecu_{\rm o}$ in $\calG_\CC(N-m+1,1)$.
As a consequence, the probability of $\vecv_{\rm s}$ decreases
monotonically with $\theta$. The larger $\theta$ is,
the smaller a ball is integrated over, the smaller are the probability
densities in the preimage integrated over, and the larger is the attenuation
$\sqrt{1-\alpha^2}$ due to the integration measure. Accordingly, the
probability distribution of $\vecv_{\rm s}$ is radial, centered at
$\vecu_{\rm s}$, and thus averaging to $\vecu_{\rm
  s}$.

To complete the proof, the distributions for
different $\eps$ have to be addressed. At $\theta=0$, the factor
$\sqrt{1-\alpha^2}$ in the  integration measure decreases with
$\eps$. Thus $f_{\Theta|E}(0,\eps)$ monotonically decreases with $\eps$. With
increasing $\theta$, the radius to integrate over in the preimage
grows with $\eps$. As a consequence, if
$f_{\Theta|E}(\theta_1,\eps_1)>f_{\Theta|E}(\theta_1,\eps_2)$ for
$\eps_1<\eps_2$, we also have
$f_{\Theta|E}(\theta_2,\eps_1)>f_{\Theta|E}(\theta_2,\eps_2)$ for
$\theta_2<\theta_1$. For $\theta>
r$, $f_{\Theta|E}(\theta,0)=0$, while for $\eps>0$, there is
probability mass in a range of $\theta>r$, according to
(\ref{eq:rperp}). 
\end{IEEEproof} \smallskip

\begin{corollary}\mabel{coro:expect}
Consider arbitrary projections $\maPi$ of a positive
semi-definite matrix $\maR$, with an ensemble of radial quantizations
$\calV$ of the eigenvectors such that the quantization $\vecv$ of
$\vecu$ is in a ball $B(\vecu)$ with radius $r\leq 1/\sqrt{2}$. The
expected Rayleigh Quotient (RQ)
$
 E_{\vecv \in B(\vecu)} \left\{{\vecv^\rmH  \maPi\maR\,\maPi \vecv }/{\vecv^\rmH\maPi\vecv} \right\}
$
is maximized for $\vecu$ the largest eigenvector $\vece$ of $\maPi\maR\maPi$.
\end{corollary}
\begin{IEEEproof}
The eigenvector $\vece$ lies in the projected space, thus
Lemma~\ref{lem:errorDistr} holds for it with $\eps=0$,
while for an arbitrary $\vecu$, generically $\eps>0$. Denote the
quantizations of these $\hat\vece$ and $\hat\vecu$. From the lemma,
the distribution of $\maPi\hat\vece$ is in a $B_e$ centered at $\vece$
with quantization error distribution $f_0(\theta)$, while
$\maPi\hat\vecu$ is in $B_u$ centered at $\maPi\vecu$ with a
larger radius and a wider distribution $f_\eps(\theta)$.

The RQ is generically non-convex, but in a ball with chordal distance
radius $1/\sqrt{2}$ around the largest eigenvector $\vece$ it is
convex. $B_e$ lies in this \columnversion{}{convexity} region.
Moreover, for any great circle going through $\vece$, the value of the
quotient on the circle at any point outside this ball is less than the
values of the circle within this ball. The same holds for the circle
arising from rotating any point in this ball in the plane spanned by
the two largest eigenvectors. Applying this in reverse order to each
eigendirection except $\vece$, one may rotate $B_u$ to be centered at
$\vece$. As the probability distribution is radial, the expectation of
the RQ does not decrease. Finally, by changing the distribution from
$f_\eps$ to $f_0$, the expectation again does not decrease due to the
great circle property.
\end{IEEEproof}

Now we are equipped for proving Proposition \ref{theo:sequential}.

\smallskip
\begin{IEEEproof}
As orthogonalization of $\maW$ does not affect its projection metric,
for OWP we compare the projection distortion of a $\maW$
created using QR-decomposition of the form (\ref{eq:wwo}) on the
feedback matrix $\maV$, resulting in an upper triangular $\maC$.

For both OWP and SWP we have orthogonal $\Nt\times K$ -matrices
$\maW$ with columns $\vecw_j$. Their   $\Nt\times j$ submatrices
are denoted by  $\maW_j =
\left[\vecw_1,
  \ldots,\vecw_j \right]$, and the corresponding projectors
(\ref{seq-orth}) to perpendicular spaces are $\maPi^\perp_j = \maI -
\maW_j\maW_j^\rmH$. The projection distortion can be decomposed as
\be
d_p(\maW,\widetilde\maR) = 1 - \sum_{j=1}^K \vecw_j^\rmH \widetilde\maR\vecw_j \,.
\ee
In both OWP and SWP, $\vecw_j$ is selected by quantizing an
eigenvector with the closest vector in a vector codebook $\calV$. In
SWP, the resulting codeword is explicitly projected by
$\maPi^\perp_{j-1}$ and normalized. In OWP with upper triangular
$\maC$, exactly the same happens. Thus both for SWP and OWP, the
projection distortion can be characterized as
\bea
\columnversion{
  d_p = 1- \sum_{j=1}^K \mu_j(\vecv_j)  \,,~~~~
 \mu_j(\vecv)  = \frac{\vecv^\rmH \maPi^\perp_{j-1} \widetilde\maR
   \maPi^\perp_{j-1}\vecv}{\vecv^\rmH \maPi^\perp_{j-1} \vecv}~, \mabel{eq:dpsum}
}{
d_p &=& 1-\sum_{j=1}^K \mu_j(\vecv_j) \mabel{eq:dpsum}\\
 \mu_j(\vecv)  &=& \frac{\vecv^\rmH \maPi^\perp_{j-1} \widetilde\maR
   \maPi^\perp_{j-1}\vecv}{\vecv^\rmH \maPi^\perp_{j-1} \vecv}~,
 \mabel{eq:RQ}
}
\eea
where $\mu_j$ is a Rayleigh Quotient (RQ) of $\widetilde\maR$ in the subspace
projected by $\maPi_{j-1}$.
For fixed $\maW_{j-1}$ and infinite granularity, the RQ
would be maximized by the largest eigenvector $\vece_{j,1}$ of
$\maPi^\perp_{j-1} \widetilde\maR \maPi^\perp_{j-1}$. With a finite granularity
codebook $\calV$, we consider three candidate codewords:
\bea
\columnversion{
\vecv_{\max} = \arg\max_{\vecv\in\calV}~\mu_j(\vecv)\,,~~~
\vecv_\perp  = \arg\min_{\vecv\in\calV}~d(\vece_{1,j},\vecv)\,,~~~
\vecv_u = \arg\min_{\vecv\in\calV}~d(\vecu_{j},\vecv)\,. \nonumber
}{
\vecv_{\max} &=& \arg\max{}_{\vecv\in\calV}~\mu_j(\vecv) \cr
\vecv_\perp  &=& \arg\min {}_{\vecv\in\calV}~d(\vece_{1,j},\vecv) \cr
\vecv_u &=& \arg\min{}_{\vecv\in\calV}~d(\vecu_{j},\vecv)\,. \nonumber
}
\eea
The first by definition minimizes the contribution of $\vecw_j$ to the projection
distortion, given the previous vectors $\maW_{j-1}$. Finding it would
require computing the RQ for each element in $\calV$.
\columnversion{Steps 7-8 in Algorithm produce $\vecv_{\perp}$}{
The second, $\vecv_{\perp}$ is the result of steps 7-8 in Algorithm
\ref{alg:seq}}, while $\vecv_u$ is the codeword selected in OWP before
orthogonalization.

By construction $\mu_j(\vecv_{\max}) \geq \mu_j(\vecv_{\perp})$ and
$\mu_j(\vecv_{u})$. From Corollary~\ref{coro:expect}, we have   
$E_{\widetilde\maR}\left\{\mu_j(\vecv_{\perp})\right\} \geq
E_{\widetilde\maR}\left\{\mu_j(\vecv_{u})\right\}$. As $\vecu$ comes from a
continuous distribution, however, equality has probability 0.

Now compare the projection distortion as a sum (\ref{eq:dpsum}) over
quantized codewords $\vecv_j$ for OWP and SWP. After $j$ terms have
been summed,  the projector for OWP is 
$\widetilde \maPi_j$ and the one for SWP $\maPi_j$. The expectation
of the remaining terms for OWP, $\sum_{k=j+1}^N\mu_k(\vecv_k)$ {\it
  does not change if} $\widetilde\maPi_j$ {\it is replaced with}
$\maPi_j$. The reason for this is that the difference between
$\maPi_j$ and $\widetilde\maPi_j$ on the remaining eigenvectors
$\vecu_k$ for $k>j$ is due to differences in quantization errors,
which produce the same result on average. Considering this after each
$j$ in the sum (\ref{eq:dpsum})  completes the proof. 
\end{IEEEproof}

\bibliographystyle{IEEEtran}	
\bibliography{CSIQuantiTrans}

\end{document}